\documentclass[12pt,pdftex]{spieman}
\usepackage{amsmath,amsfonts,amssymb}
\usepackage{graphicx}
\usepackage{setspace}
\usepackage{tocloft}
\usepackage{siunitx,diffcoeff}
\usepackage{lipsum}
\usepackage[right]{lineno}
\usepackage[capitalise]{cleveref}
\usepackage{booktabs}
\usepackage{multirow}
\usepackage{xspace}
\usepackage[dvipsnames]{xcolor}

\usepackage[T1]{fontenc}
\usepackage{mathptmx,tgheros,inconsolata}

\DeclareMathAlphabet{\mathcal}{OMS}{cmsy}{m}{n}
\DeclareSIUnit\pixel{pix}
\DeclareSIUnit\electron{e^{-}}
\DeclareSIUnit\textarcsec{as}
\DeclareSIUnit\mas{\milli\textarcsec}
\DeclareSIUnit\uas{\micro\textarcsec}
\DeclareSIUnit\micron{\micro\meter}
\DeclareSIUnit\parsec{pc}
\DeclareSIUnit\kpc{\kilo\parsec}
\DeclareSIUnit\Rsun{R_\odot}
\DeclareSIUnit\Lsun{L_\odot}

\newcommand{\plateanalysis}{{Plate Analysis}\xspace}
\newcommand{\gaia}{{\itshape Gaia}\xspace}

\title{Demonstration of Plate Analysis, an algorithm for precise relative astrometry}

\author[a]{Ryou Ohsawa}
\author[a,b]{Daisuke Kawata}
\author[c]{Takafumi Kamizuka}
\author[d]{Yoshiyuki Yamada}
\author[e]{Wolfgang L\"offler}
\author[e]{Michael Biermann}
\affil[a]{JASMINE project, National Astronomical Observatory of Japan, Japan}
\affil[b]{Mullard Space Science Laboratory, University College London, United Kingdom}
\affil[c]{Institute of Astronomy, Graduate School of Science, The University of Tokyo, Japan}
\affil[d]{Theoretical Astrophysics Group, Department of Physics, Kyoto University, Japan}
\affil[e]{Astronomisches Rechen-Institut, Zentrum f\"ur Astronomie der Universit\"at Heidelberg, Germany}

\cftpagenumbersoff{figure}
\cftpagenumbersoff{table}

\begin{document}
\maketitle

\begin{abstract}
Astrometry plays a crucial role in understanding the structure, dynamics, and evolution of celestial objects by providing precise measurements of their positions and motions. We propose a new approach to wide-field, relative astrometry, Plate Analysis. Plate Analysis is an innovative algorithm that estimates stellar coordinates and corrects geometric distortion based on precise reference sources, without relying on dedicated calibration fields. It is implemented as a probabilistic framework using Stochastic Variational Inference to efficiently optimize the numerous parameters involved in the model.

The methodology was tested through a simplified simulation designed after the Galactic center survey of the JASMINE mission. This simulation, called the JASMINE mini-mock survey, covered three years of observation with 100 satellite orbits, providing a comprehensive dataset for evaluating the performance of Plate Analysis. Although the observation model incorporated more than 30,000 parameters, the parameters were efficiently optimized through Plate Analysis. The results showed that the estimated coordinates closely match the expected stellar motions, with an average positional error of about \qty{70}{\uas}. These findings validate the potential of Plate Analysis for precise wide-field astrometric applications, offering significant insights into improving the accuracy of stellar dynamics measurements.
\end{abstract}


{\noindent \footnotesize\textbf{*}Ryou Ohsawa, \linkable{ryou.ohsawa@nao.ac.jp} }
\section{Introduction}
\label{sec:intro}

The formation process of the Milky Way remains a big problem in astronomy. The kinematics and metallicities of stars is a promising clue to solving the history of the Milky Way.\cite{eggen_evidence_1962,chiba_kinematics_2000} However, further studies require ultimate precision in astrometry, the measurements of stellar positions. Space missions have largely cultivated precise astrometry, Hipparcos\cite{perryman_hipparcos_1997} and \gaia\cite{gaia_collaboration_gaia_2016}. Hipparcos provided astrometric catalogs of about 1.2 million nearby stars with a mas-level precsion\cite{perryman_hipparcos_1997,van_leeuwen_validation_2007}. The precise Hipparcos data revealed a blob of halo stars in the angular momentum space\cite{helmi_debris_1999}, which is possibly a remnant of a dwarf galaxy that was incorporated in the Milky Way. This scenario was supported by N-body simulations\cite{brook_galactic_2003}. Dedicated photometric and spectroscopic studies have also contributed to establishing the Milky Way formation scenario, elucidating that the Galactic halo contains multiple stellar populations with different kinematics.\cite{yanny_identification_2000,newberg_ghost_2002,belokurov_field_2006,carollo_two_2008,roederer_chemical_2009,ishigaki_chemical_2010,ishigaki_erratum_2010,nissen_two_2010} Then, \gaia, the successor of Hipparcos, measured about 2 billion stars and released a sequence of catalogs with a {\textmu}as-level precision. A spiral feature was identified in the position-velocity space of disk stars (phase spiral)\cite{antoja_dynamically_2018,antoja_phase_2023}. Several studies\cite{binney_origin_2018,bland-hawthorn_galactic_2021,antoja_phase_2023} suggested that the phase spiral was the wake of the passage of a dwarf galaxy about 1 Gyr ago. \gaia also discovered that some parts of halo stars exhibited unusually large radial velocities\cite{belokurov_co-formation_2018}, possibly a sign of a major merger occurred about 10 Gyr ago\cite{villalobos_simulations_2008,villalobos_simulations_2009}. The merger event with a massive satellite galaxy, named \textit{Gaia-Sausage-Enceladus}, formed the inner halo and the thick disk of Milky Way\cite{belokurov_co-formation_2018,vincenzo_fall_2019,grand_dual_2020}. These discoveries strongly support the scenario where the Milky Way has grown through many mergers with satellite galaxies.

Hipparcos and \gaia successfully demonstrated the power of precise astrometry in studying the Milky Way formation. However, they cannot inspect the Galactic center region as long as they observe in the optical wavelengths. Visible photons are easily absorbed by interstellar dust. Thus, \gaia can observe stars within about \qty{4}{kpc} toward the Galactic center region. Bulge stars near the Galactic disk are rarely observed by \gaia. Recent ground-based infrared observations revealed an inner structure within the bulge. A stellar disk spreading about 200 parsecs (Nuclear Stellar Disk, NSD) resides within the Galactic center and is associated with a dense molecular cloud (Central Molecular Zone, CMZ). There is a cluster (Nuclear Stellar Cluster, NSC) at the center of the NSD\cite{chen_star_2023}. Although the Galactic center region has been intensively investigated in the infrared wavelengths\cite{nogueras-lara_galacticnucleus_2018,nogueras-lara_galacticnucleus_2019}, the formation processes of the NSD and NSC remain a matter of debate\cite{matsunaga_lack_2016,nogueras-lara_star_2018,nogueras-lara_nuclear_2021}. There is a great demand for precise infrared astrometry, which may provide the kinematics of stars around the Galactic center region.

Hipparcos and \gaia are facilities specially designed to scan the whole sky to define the reference system. Their approach is characterized as global, absolute astrometry. On the other hand, standard observatory-type telescopes can precisely measure the relative positions of sources in a small region. For example, the supermassive black hole at the Galactic center, \textit{Sagittarius} A$^*$, was identified by the intensive monitoring with Keck and NTT\cite{eckart_observations_1996,genzel_nature_1997,ghez_high_1998,ghez_accelerations_2000,eckart_stellar_2002,schodel_star_2002}. These works are characterized as local, relative astrometry.

Sometimes the targets of interest can be extended beyond the telescope's field of view, in cases of stellar clusters\cite{anderson_new_2010}, the Magellanic clouds\cite{niederhofer_hubble_2024}, or the nuclear stellar disk\cite{shahzamanian_first_2019,shahzamanian_proper_2022}. These kinds of investigations require the wide-field, relative astrometry, where a geometric distortion of obtained images is sufficiently corrected, and the stellar positions are adequately mapped on the reference frame. A prescription was proposed based on observations with the Hubble Space Telescope\cite{anderson_improved_2003}. Updated variants have been applied to several ground-based observations\cite{anderson_ground-based_2006,yadav_ground-based_2008,bellini_ground-based_2009,bellini_ground-based_2010,libralato_ground-based_2014}. As part of the VISTA Variables in the Via Lactea (VVV) survey\cite{saito_milky_2012,soto_milky_2013,minniti_milky_2014,alonso-garcia_milky_2018}, the VISTA telescope\cite{dalton_vista_2006} repeatedly observed the Galactic center region and produced astrometric catalogs\cite{contreras_ramos_proper_2017,smith_virac_2018,libralato_high-precision_2015,griggio_high-precision_2024}. Recently, ESA's Euclid\cite{euclid_collaboration_euclid_2022} starts observations, and its superb image quality enables high-precision astrometry\cite{libralato_euclid_2024}.

In wide-field relative astrometry, geometric distortion is typically estimated using a dedicated calibration field, assuming that this distortion remains constant throughout the observation period. However, in some practical situations, this assumption proves to be overly simplistic. For instance, when a satellite telescope is in low-Earth orbit, it is subject to variable irradiation by the Earth from different angles, which causes thermal deformation of the optics. This results in slight but significant changes to the geometric distortion over time. Frequent observations of the calibration field to adjust for these changes are not always feasible.

To address these challenges, we propose a new approach called ``\plateanalysis''. Unlike traditional methods, \plateanalysis does not require the calibration field and estimates the geometric distortion directly from the set of plates (exposures) for science, leveraging a few reference anchor points as guides. The new observation model is implemented within a probabilistic modeling framework. By utilizing a differentiable programming scheme, we efficiently obtain the astrometric solution, and the positions and uncertainties of sources, making the approach computationally viable and well-suited for advanced astronomical applications.

This paper presents the definition of \plateanalysis and its implementation. The performance of the algorithm is evaluated based on a simplified and small-scale survey designed after the JASMINE mission (hereafter, mini-mock survey), where we analyze the exposures obtained in each orbit separately and demonstrate that the apparent motions of stars are successfully reproduced. The paper is organized as follows. The definition of the mini-mock survey and the observation model are presented in \cref{sec:method}. The details of the survey simulation are described in \cref{sec:experiment}. The results are illustrated in \cref{sec:results}. The performances of the algorithm are discussed in \cref{sec:discussion}. Then, \cref{sec:conclusion} provides the conclusion.
\section{Method}
\label{sec:method}

\subsection{Plate Analysis}
\label{sec:exp:plate}

Accurately measuring the coordinates of sources in a certain region is an important step in wide-field, relative astrometry. However, astrometric measurements may suffer from several disturbances, such as thermal deformation of optics, satellite jitter motions, and aberration. Correcting the geometric distortion is of great importance in precise astrometry. Evaluating the magnitude of these disturbances in advance is not realistic. Thus, they should be corrected based on the data themselves, i.e. self-calibration. A feasible method for self-calibration is estimating the geometric distortion in repeated observations of a calibration field\cite{anderson_improved_2003}.

Nevertheless, in case that the geometric distortion is not fully stable with time, frequent calibrations are required, deteriorating the observation efficiency. Another possible method is optimizing an observation model that incorporates the celestial sources, the satellite systems, and every noise source. This approach has been used in global, absolute astrometry. In the Hipparcos mission, measurements obtained in each great circle were analyzed at once\cite{perryman_hipparcos_1997,van_leeuwen_validation_2007}. Instead, \gaia uses an algorithm that can handle all the measurements at once\cite{lindegren_astrometric_2012,lindegren_gaia_2018,lindegren_gaia_2021}, leading to a global astrometric solution.

We propose an approach to achieve wide-field, relative astrometry. The measurements are separated by epochs and analyzed independently, like the Hipparcos scheme. We estimate the coordinates of sources at the observation epoch using a set of measurements, assuming that the stellar motions are negligible during the observation period. We also assume that the change in image distortion (geometric distortion) is negligible during the observation period. The image distortion is estimated based on a set of science frames and some reference sources as anchor points. We call this scheme ``\plateanalysis''. The implementation of the observation model in this paper is defined in the following section.

\subsection{Observation model}
\label{sec:method:model}

The observation model defines the mapping from the spherical, celestial coordinates to the two-dimensional detector coordinates. This section describes the sequence of coordinate transformations in the observation model.

The apparent positions at the observation epoch are the parameters of interest, denoted by $(\alpha^\text{src}_{i}, \delta^\text{src}_{i})$, where $i$ is the unique source index. Then, define the telescope pointing direction as $(\alpha^\text{tel}_{m}, \delta^\text{tel}_{m})$, where $m$ is the unique exposure index. The separation angle between the source and the telescope pointing, $r_{i,m}$, is given by
\begin{equation}
\cos{r_{i,m}} = \sin{\delta^\text{tel}_{m}} \sin{\delta^\text{src}_{i}}
+ \cos{\delta^\text{tel}} \cos{\delta^\text{src}_{i}}
\cdot \cos(\alpha^\text{src}_{i} - \alpha^\text{tel}_{m}).
\label{eq:method:cosr}
\end{equation}
They are projected onto an idealized focal plane centered at the telescope pointing direction. Here, we use the Gnomonic projection as a simple case. The intermediate coordinates $(\xi_{i,m}, \eta_{i,m})$, which are usually referred as to the Standard coordinates\cite{green_spherical_1985}, are obtained below.
\begin{equation}
\left\{~
\begin{aligned}
\xi_{i,m}
& = {\frac{\cos{\delta^\text{src}_{i}}
\sin(\alpha^\text{src}_{i} - \alpha^\text{tel}_{m})}{\cos{r_{i,m}}}} \\
\eta_{i,m}
& = {\frac{\sin{\delta^\text{src}_{i}} \cos{\delta^\text{tel}_{m}}
- \sin{\delta^\text{tel}_{m}} \cos{\delta^\text{src}_{i}}
\cos(\alpha^\text{src}_{i} - \alpha^\text{tel}_{m})}{\cos{r_{i,m}}}}
\end{aligned}
\right.
\label{eq:method:projection}
\end{equation}
Then, the intermediate coordinates are rotated by the telescope position angle $\theta^\text{tel}_{m}$ and scaled by the telescope focal length. We obtain the coordinates on the idealized focal plane, $(X_{i,m}, Y_{i,m})$.
\begin{equation}
\begin{pmatrix}
X_{i,m} \\
Y_{i,m}
\end{pmatrix}
=
F_{m}
\begin{pmatrix}
\cos{\theta^\text{tel}_{m}} & \sin{\theta^\text{tel}_{m}} \\
-\sin{\theta^\text{tel}_{m}} & \cos{\theta^\text{tel}_{m}}
\end{pmatrix}
\begin{pmatrix}
{-}\xi_{i,m} \\
{+}\eta_{i,m}
\end{pmatrix},
\label{eq:method:ideal}
\end{equation}
where $F_{m}$ is the focal plane scaling factor of the $m$th exposure.

Since images suffer from distortion, the positions on the actual focal plane can be displaced from the idealized cases. We assume that the displacement can be described using analytical functions. Here, we defined the displacements using the Legendre polynomials of up to $N_\mathcal{L}$-th order.
\begin{align}
\hat{X}_{i,m} &
= X_{i,m} + \sum_{k+l \le N_\mathcal{L}} A_{k,l}
\mathcal{L}_k(s\, X_{i,m}) \mathcal{L}_l(s\, Y_{i,m}),
\label{eq:model:Lx} \\
\hat{Y}_{i,m} &
= Y_{i,m} + \sum_{k+l \le N_\mathcal{L}} B_{k,l}
\mathcal{L}_k(s\, X_{i,m}) \mathcal{L}_l(s\, Y_{i,m}).
\label{eq:model:Ly}
\end{align}
The function $\mathcal{L}_k(\cdot)$ is the $k$th-order Legendre polynomial, and $A_{k,l}$ and $B_{k,l}$ are the coefficients of the polynomials. The maximum order $N_\mathcal{L}$ is tentatively set 5, which is usually sufficient to represent ordinal optical distortion The fixed factor $s$ is applied to $X_{i,m}$ and $Y_{i,m}$ so that $s\,{X_{i,m}} \in [-1, 1]$ and $s\,{Y_{i,m}} \in [-1, 1]$. The uniform displacements have virtually the same effect as the offsets of the telescope pointing. The coefficients of the 0th-order term are constrained to eliminate the displacement of the optical center.
\begin{align}
A_{0,0} & = \frac{A_{2,0} + A_{0,2}}{2}
- \frac{A_{2,2}}{4} - \frac{3 \left( A_{4,0} + A_{0,4} \right)}{8},
\label{eq:model:A0} \\
B_{0,0} & = \frac{B_{2,0} + B_{0,2}}{2}
- \frac{B_{2,2}}{4} - \frac{3 \left( B_{4,0} + B_{0,4} \right)}{8}.
\label{eq:model:B0}
\end{align}
The first-order terms can degenerate with several telescope parameters (focal lengths and position angle). The coefficients of the first-order terms are tentatively set to zeros to avoid a possible degeneracy.
\begin{equation}
A_{1,0} = 0, ~~ A_{0,1} = 0, ~~ B_{1,0} = 0,~~ B_{0,1} = 0.
\label{eq:model:first}
\end{equation}
Thus, the distortion of the telescope is characterized by the 18 coefficients in each direction.

Assume that a source falls onto the $n$th detector. An affine transformation is used to convert the focal plane coordinates $(\hat{X}_{i,m}, \hat{Y}_{i,m})$ to the detector coordinates $(\hat{n}_{x,i,m}, \hat{n}_{y,i,m})$.
\begin{equation}
\begin{pmatrix}
\hat{n}_{x,i,m} \\
\hat{n}_{y,i,m}
\end{pmatrix}
=
\begin{pmatrix}
\left({\delta^\text{det}_{x,n}}\right)^{-1} & 0 \\
0 & \left({\delta^\text{det}_{y,n}}\right)^{-1}
\end{pmatrix}
\begin{pmatrix}
\cos\theta^\text{det}_n &  -\sin\theta^\text{det}_n \\
\sin\theta^\text{det}_n & \cos\theta^\text{det}_n
\end{pmatrix}
\begin{pmatrix}
\hat{X}_{i,m} - X^\text{det}_n \\
\hat{Y}_{i,m} - Y^\text{det}_n
\end{pmatrix},
\label{eq:exp:model:nxny}
\end{equation}
where $X^\text{det}_n$ and $Y^\text{det}_n$ are the origins of the detector coordinate on the focal plane, $\theta^\text{det}_n$ is the rotation angle of the detector, and $\delta^\text{det}_{x,n}$ and $\delta^\text{det}_{y,n}$ are the pixel sizes. The subscript $n$ denotes the detector number. The sources that fall outside the detectors are ignored.

\subsection{Implementation \& Optimization}
\label{sec:method:optim}

We implemented the observation model using \texttt{JAX}\cite{bradbury_jax_2018}, which is a powerful differentiable programming framework in Python. Thus, all the transformations were implemented as differentiable. The observation model was handled under a probabilistic programming framework using \texttt{numpyro}\cite{phan_composable_2019}. We used the variational inference to optimize the model to simultaneously estimate the parameters and their uncertainties. Thanks to \texttt{JAX} and \texttt{numpyro}, we were able to construct the observation model as a probabilistic model without any difficulties. Details of the parameter inference are described below.

We define $\mathcal{D} = \{ (\hat{n}_{x,i,m}, \hat{n}_{y,i,m}), \ldots \}$ as a set of measurements. From the Bayes theorem, the posterior probability density function of the celestial coordinates $({\alpha}^\text{src}_{i}, {\delta}^\text{src}_{i})$ is provided by the product of the likelihood and the priors.
\begin{equation}
\mathcal{P}(
{\alpha}^\text{src}_{i}, {\delta}^\text{src}_{i} ~|~
\mathcal{D}
)
\propto
\underset{\text{observation model}}{\underline{
\mathcal{P}(
\mathcal{D} ~|~
{\alpha}^\text{src}_{i}, {\delta}^\text{src}_{i},
\ldots
)}}~~
\underset{\text{priors}}{\underline{
\mathcal{P}(
{\alpha}^\text{src}_{i}, {\delta}^\text{src}_{i},
\ldots
)}},
\label{eq:method:posterior:radec}
\end{equation}
where $\mathcal{P}(\mathcal{X} ~|~ \cdots)$ denotes a probability density function of $\mathcal{X}$ with conditions. The observation model defines the likelihood. For simplicity, we assume that the measurements follow the independent normal distributions. The prior term plays an important role in \plateanalysis. Relative astrometry requires some reference stars as anchor points. The positions and uncertainties of the reference sources are naturally introduced as the priors of \cref{eq:method:posterior:radec}. We can also adopt strong priors for non-astrometric parameters to mitigate possible parameter degeneracy. In this paper, we use the normal distributions to describe the priors.

Although we can obtain the posterior distributions by sampling from \cref{eq:method:posterior:radec}, to accelerate the inference, we aggressively approximate the posterior by the product of the normal distributions.
\begin{equation}
\mathcal{P}(
\mathcal{D} ~|~
{\alpha}^\text{src}_{i}, {\delta}^\text{src}_{i},
\ldots
)~
\mathcal{P}(
{\alpha}^\text{src}_{i}, {\delta}^\text{src}_{i},
\ldots
)
\simeq
\mathcal{N}(
{\alpha}^\text{src}_{i} ~|~
\hat{\alpha}^\text{src}_{i}, \sigma^\text{src}_{\alpha, i}) ~
\mathcal{N}(
{\delta}^\text{src}_{i} ~|~
\hat{\delta}^\text{src}_{i}, \sigma^\text{src}_{\delta, i}),
\label{eq:method:posterior:svi}
\end{equation}
where $\mathcal{N}(\mu, \sigma)$ denotes the normal distribution whose mean value and standard deviation are $\mu$ and $\sigma$, respectively. The parameters of the right-hand side, $\hat{\alpha}^\text{src}_{i}, \sigma^\text{src}_{\alpha, i}, \hat{\delta}^\text{src}_{i}, \sigma^\text{src}_{\delta, i}$, are optimized to minimize the Kullback-Leibler divergence between both sides. This procedure is usually referred to the Stochastic Variational Inference (SVI) \cite{hoffman_stochastic_2013}. The probabilistic modeling was implemented using \texttt{numpyro}.

The SVI optimization is carried out with {ADAM}\cite{kingma_adam_2015}. {ADAM} is a robust optimizer that takes the advantages of {RMSProp}\cite{hinton_lecture_2012} and {Momentum}\cite{polyak_methods_1964}. ADAM has four parameters $\alpha$, $\epsilon$, $\beta_1$, and $\beta_2$. To stabilize the parameter optimization, we gradually reduced the learning rate $\alpha$ from $10^{-3}$ to $10^{-10}$. The other parameters are listed in Table\,\ref{tab:exp:optim:adam}.

\begin{table}[t]
\centering
\caption{Parameters of the ADAM optimizer}
\label{tab:exp:optim:adam}
\begin{tabular}{lcl}
\toprule
\multicolumn{1}{c}{Name}
& \multicolumn{1}{c}{Value}
& \multicolumn{1}{c}{Comment} \\
\midrule
$\alpha$ & $10^{-3}$--$10^{-10}$
& Learning rate of the optimization \\
$\epsilon$ & $10^{-4}$
& Regularization factor to avoid the zero division \\
$\beta_1$ & 0.99
& Momentum factor \\
$\beta_2$ & 0.999
& Gradient normalizing factor \\
\bottomrule
\end{tabular}
\end{table}
\section{Experiment}
\label{sec:experiment}

We validated the proposed algorithm based on numerical experiments. Mock data for an input catalog and corresponding observations were generated along with a simulated survey observation. The coordinates at the observation periods are estimated individually. Then, we compare the time sequence of the estimated celestial coordinates with ground truth. In this paper, we designed a small and simplified survey after the Galactic center survey of the JASMINE mission.

\subsection{JASMINE mission}
\label{sec:exp:jasmine}

To illuminate the mystery of the Galactic center, we initiated the Japan Astrometry Satellite Mission for Infrared Exploration (JASMINE) project\cite{gouda_infrared_2019,kawata_jasmine_2024}. JASMINE is designed to achieve supreme accuracies in astrometry and photometry in the near-infrared wavelength (\qtyrange{1.0}{1.6}{\micro\meter}). The conceptual design of the JASMINE payload is presented by Kataza et al. (2024)\cite{kataza_conceptual_2024}. JASMINE will be launched into a Sun-synchronous (dawn/dusk) orbit at an altitude of about \qty{600}{\kilo\meter}. JASMINE is equipped with a telescope with a diameter of \qty{36}{\centi\meter}, made from materials with extremely low thermal expansion coefficients (${<}\qty{1e-8}{\kelvin^{-1}}$)\cite{isobe_structural_2024}. A sophisticated heater control enables JASMINE to keep the temperature variation of the optics within $\Delta{T} \lesssim \qty{0.1}{\kelvin}$ during an orbit, even in the low Earth orbit (LEO). The structure and performance of the JASMINE optics are discussed in Suematsu et al.\cite{suematsu_evaluation_2024} and Isobe et al.\cite{isobe_structural_2024}. The camera covers about a $\ang{0.55}\times\ang{0.55}$ region, equipped with four InGaAs image sensors developed by Hamamatsu Photonics\cite{nakaya_low_2016,miyakawa_performance_2024}. Each sensor has $1968{\times}1968$ pixels in a \qty{10}{\micron} interval, and the pixel scale is about $\ang{;;0.472}{\times}\ang{;;0.472}$ in the sky.

The science field of the JASMINE Galactic center survey is $\ang{2.1}{\times}\ang{1.2}$ in Galactic coordinates. While Hipparcos and \gaia scanned the sky while spinning around their rotation axes, JASMINE will repeatedly observe the Galactic center region in a step-stare method. Thus, JAMSINE's observation strategy is quite different from those of Hipparcos and \gaia. Since the science field does not contain bright quasars, JASMINE cannot define its own reference frame and relies on bright foreground stars measured by \gaia to determine the absolute positions and scales of the obtained images. Hubble Space Telescope has successfully measured proper motions and parallaxes with supreme accuracy comparable to \gaia.\cite{casertano_parallax_2016,riess_new_2018,libralato_hubble_2023} Recently, precise infrared astrometric measurements are achieved with JWST and Euclid.\cite{libralato_jwst-tst_2023,libralato_high-precision_2024,libralato_euclid_2024}. Shortly, Roman Space Telescope will provide accurate positional measurements in infrared.\cite{terry_galactic_2023} Although JASMINE is a small telescope compared to those space telescopes, about half the mission period will be dedicated to the astrometry mission, and the specifications are optimized for bright red giant stars. JASMINE will be a competitive mission in near-infrared, wide-field, and relative astrometry.\cite{kawata_jasmine_2024}

\subsection{JASMINE mini-mock survey}
\label{sec:exp:survey}

We generated mock measurements for a simplified and small-scale survey called the JASMINE mini-mock survey. This section describes the observation design and schedule. The data used in the experiment is publicly hosted in Zenodo (\href{https://zenodo.org/records/10403895}{Version 20240602})\cite{ohsawa_jasmine_2024}.

\subsubsection{Survey Design}
\label{sec:exp:survey:design}

The JASMINE Galactic center science field is defined in Galactic coordinates as $\ang{-1.4} \le \ell \le \ang{0.7}, \ang{-0.6} \le b \le \ang{0.6}$. The actual observation field has a peripheral margin of \ang{0.55}, which is called the JASMINE Galactic center extended science field. JASMINE will repeatedly observe the entire science field, but we do not need to cover the entire field for the current purpose. Thus, we defined the target coordinate $(\ell, b) = (\ang{-0.3}, \ang{0.1})$ and arranged the observation schedule so that the target coordinate was intensively observed.

JASMINE can observe the Galactic center region only in the spring and autumn seasons. The observation schedule is summarized in \cref{tab:exp:epoch}. The operation covers about three years with six observation periods. Since JASMINE has a sun shield in a single direction\cite{kataza_conceptual_2024}, the telescope position angle is different by about \ang{180} between the spring and autumn seasons. JASMINE will observe the Galactic center region for more than 6,000 orbits, but we limited the number of orbits to 100 for the simplicity of simulation. The observation starting time of each orbit was defined by uniformly splitting the observation periods.

\begin{table}[t]
\centering
\caption{Definition of the observation periods}
\label{tab:exp:epoch}
\begin{tabular}{ccc}
\toprule
& \multicolumn{1}{c}{Spring}
& \multicolumn{1}{c}{Autumn} \\
\midrule
Year 1 & 2028-02-01 to 2028-05-01 & 2028-08-01 to 2028-11-01 \\
Year 2 & 2029-02-01 to 2029-05-01 & 2029-08-01 to 2029-11-01 \\
Year 3 & 2030-02-01 to 2030-05-01 & 2030-08-01 to 2030-11-01 \\
\bottomrule
\end{tabular}
\end{table}

JASMINE has a focal plane of about $41{\times}\qty{41}{mm}$, corresponding to a \ang{0.55} square in the sky. In this experiment, the telescope has a slightly smaller field of view ($\ang{0.48}{\times}\ang{0.48}$). The origin of the telescope position angle is defined so that the $y$-axis of the focal plane is roughly aligned parallel to the Galactic latitude axis. The telescope position angle was set about \ang{0} and \ang{180} in the spring and autumn seasons, respectively.

Evaluating the distortion patterns in advance is practically impossible. The distortion should be estimated by self-calibration using the stellar positions measured on the detectors. In Plate Analysis, it is important to measure sources in different positions on the focal plane during a single satellite orbit. Thus, we tentatively adopted a mapping scheme in which two pairs of fields were observed in a single orbit. The fields of each pair overlapped each other by half. One pair was aligned in the Galactic latitude direction, and the other was aligned in the Galactic longitude direction. The factors that describe the observation schedule are summarized in \cref{tab:exp:schedule}.

\begin{table}
\centering
\caption{Specifications of Survey Schedule}
\label{tab:exp:schedule}
\begin{tabular}{>{\ttfamily}lcl}
\toprule
\multicolumn{1}{c}{Name}
& \multicolumn{1}{c}{Symbol}
& \multicolumn{1}{c}{Comment} \\
\midrule
orbit\_id & $N_\text{orbit}$
& Unique Orbit ID, 0-origin, ranging from 0 to 99 \\
field\_id & $N_\text{field}$
& Field ID, 0-origin, ranging from 0 to 3 \\
plate\_id & $N_\text{plate}$
& Exposure ID, 0-origin, ranging from 0 to 23 \\
glon & $\ell^\text{tel}_{0}$
& Longitudes in Galactic coordinates \\
glat & $b^\text{tel}_{0}$
& Latitudes in Galactic coordinates' \\
pa & $\theta^\text{tel}_{0}$
& Position angles in Galactic coordinates \\
obstime & $T_\text{obs}$
& Timestamp of each plate as the MJD format in the TCB scale \\
\bottomrule
\end{tabular}
\end{table}

The telescope pointing will be subject to errors. We define the scheduled pointing direction as $(\ell^\text{tel}_{0}, b^\text{tel}_{0}, \theta^\text{tel}_{0})$. The following rules provided the actual pointing direction.
\begin{equation}
\ell^\text{tel}_{1}
= \mathcal{N}(\ell^\text{tel}_{0}, \sigma_{\ell, \text{tel}, 0}), \quad
b^\text{tel}_{1}
= \mathcal{N}(b^\text{tel}_{0}, \sigma_{b, \text{tel}, 0}), \quad
\theta^\text{tel}_{1}
= \mathcal{N}(\theta^\text{tel}_{0}, \sigma_{\theta, \text{tel}, 0}),
\label{eq:exp:pointing}
\end{equation}
where $\sigma_{\ell, \text{tel}, 0} = \qty{1.0}{arcsec}/\cos(b^\text{tel}_{0})$, $\sigma_{b, \text{tel}, 0} = \qty{1.0}{arcsec}$, and $\sigma_{\theta, \text{tel}, 0} = \qty{1.0}{arcsec}$. In each field, 24 exposures were successively obtained. The telescope direction can be altered by jittering due to the telescope pointing errors. The telescope pointing of each exposure was provided by the following rules.
\begin{equation}
\ell^\text{tel}_{m}
= \mathcal{N}(\ell^\text{tel}_{m-1}, \sigma_{\ell, \text{tel}}), \quad
b^\text{tel}_{m}
= \mathcal{N}(b^\text{tel}_{m-1}, \sigma_{b, \text{tel}}), \quad
\theta^\text{tel}_{m}
= \mathcal{N}(\theta^\text{tel}_{m-1}, \sigma_{\theta, \text{tel}}),
\label{eq:exp:pointing:succeed}
\end{equation}
where $\sigma_{\ell, \text{tel}} = \qty{1.0}{arcsec}/\cos(b^\text{tel}_{0})$, $\sigma_{b, \text{tel}} = \qty{1.0}{arcsec}$, and $\sigma_{\theta, \text{tel}} = \qty{1.0}{arcsec}$. We assumed that it took \qty{90}{sec} for changing fields and \qty{20}{sec} for each exposure.

Since JASMINE orbits in the LEO, the satellite temperature can gradually change due to the changing irradiation from the Earth. Thus, the focal length can change with every exposure. The actual focal lengths were defined by the following rules:
\begin{equation}
F_{n} = F_{0}\cdot s_{n}, \quad
s_{0} \sim \mathcal{N}(1, \sigma_{F_0}), \quad
s_{n}/s_{n-1} \sim \mathcal{N}(1, \sigma_{F}),
\label{eq:exp:efl}
\end{equation}
where $F_{0}$ is the nominal focal length. We adopted $\sigma_{F_0} = 0.001$ and $\sigma_{F} = 0.0001$.

JASMINE utilizes thermally stable optics\cite{kataza_conceptual_2024,isobe_structural_2024}. We expect that the image distortion pattern does not change within a satellite orbit. In this experiment, we assumed that the distortion pattern was constant during the mission, and the data obtained in a single orbit were independently analyzed in \plateanalysis. The coefficients were set as zeros except for the following terms:
\begin{equation}
\begin{aligned}
A_{1,1} &= \phantom{-}0.2,
& A_{3,0} &= \phantom{-}3.0,
& A_{5,0} &= \phantom{-}0.3, \\
B_{1,1} &= \phantom{-}0.3,
& B_{0,3} &= -3.0,
& B_{0,5} &= \phantom{-}0.5.
\end{aligned}
\label{eq:exp:distortion}
\end{equation}
Those coefficients were tentatively selected to represent a moderate distortion pattern caused by optics misalignment or deformation. The amplitudes of the coefficients were assigned so that the distortion can cause displacements as large as \qty{0.01}{pixel} ($= \qty{4}{\mas}$).

The focal plane has 4 detectors aligned in a $2{\times}2$ grid with \qty{3}{mm} gaps. Each detector has $1920{\times}1920$ pixels with \qty{10}{\micro\meter} sides. We assumed that the alignments of the detectors were constant during the mission. In the optimization process, the detector positions and the pixel scales were regarded as unknown parameters.

\subsubsection{Ground truth catalog}
\label{sec:exp:survey:gt}
We defined the ground truth catalog to generate mock observations. Since the simulation requires an astrometric catalog with proper motions and parallaxes, the \gaia DR3 catalog was used as the ground truth. We did not include infrared observations for the sake of simplicity. Thus, all the stars were observed in the optical wavelengths, and the stellar distribution differed from that in the infrared. The column definitions of the ground truth catalog are summarized in \cref{tab:exp:catalog}.

To regulate the number of sources, we selected the stars whose parallaxes were measured with $S/N > 5$ in the JASMINE extended science field from the \gaia DR3 catalog. The number of the extracted sources was 132,651. Since the sources were selected by the parallax accuracies, the ground truth catalog contains many nearby sources with large parallaxes.

\begin{figure}[t]
\centering
\includegraphics[width=0.85\linewidth]{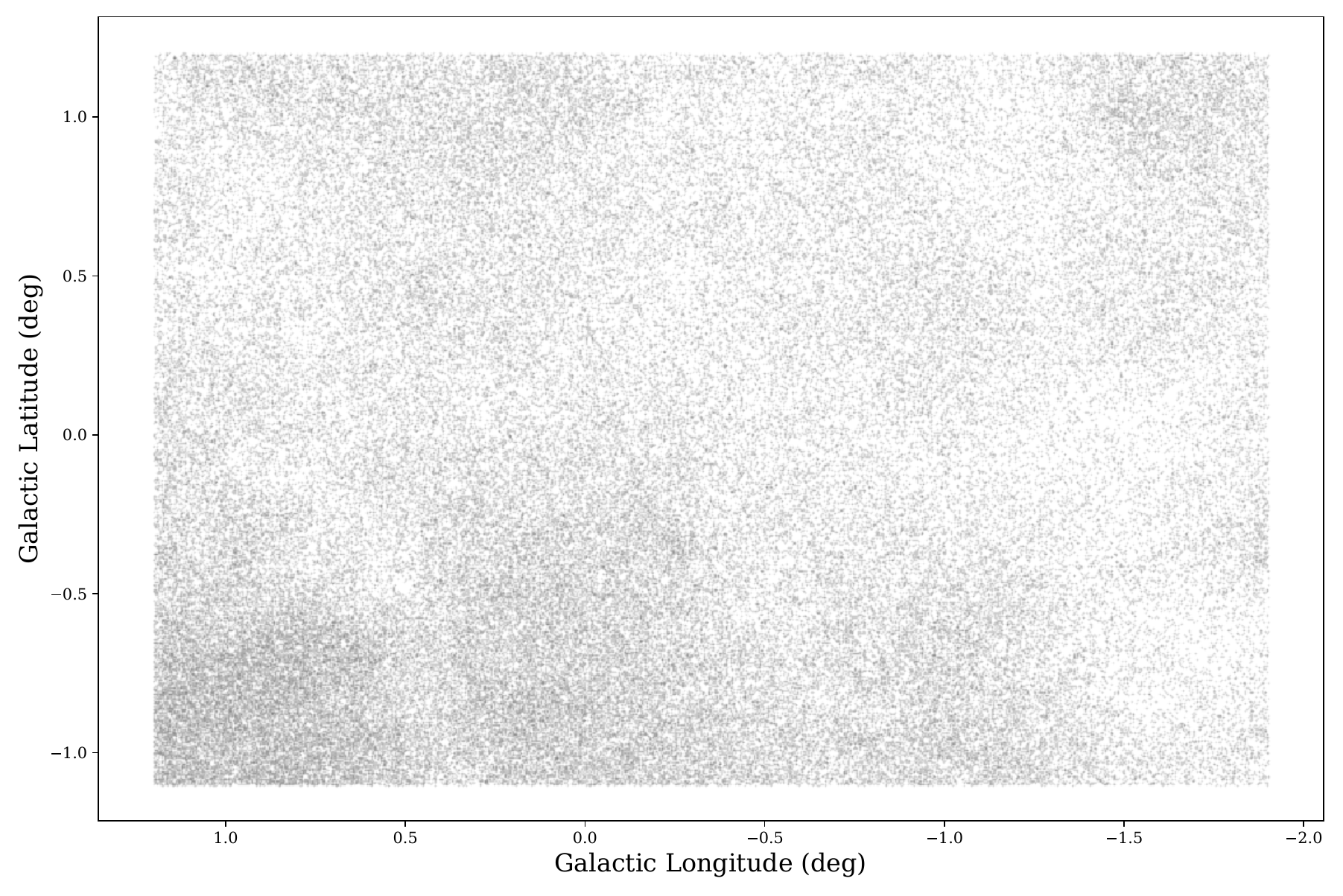}
\caption{Distribution of the sources in the ground truth catalog. The sources selected from the \gaia DR3 catalog are illustrated in Galactic coordinates.}
\label{fig:exp:source}
\end{figure}

To calibrate the measurements to ICRS, JASMINE has to rely on external information. Currently, we plan to use the \gaia sources whose astrometric parameters are tightly constrained. We selected reference sources such that the uncertainties of the parallaxes $\sigma_\varpi < \qty{25}{\uas}$ and the positional uncertainties at the observation epoch $\sigma_\Delta < \qty{500}{\uas}$. The positional error $\sigma_\Delta$ is defined below.
\begin{equation}
\sigma_\Delta^2 =
\sigma_{\alpha,\text{total}}^2 + \sigma_{\delta,\text{total}}^2
~~ \left\{
\begin{aligned}
\sigma_{\alpha,\text{total}}^2
&= \sigma_{\alpha}^2 +
(\Delta{t}\, \sigma_{\mu_{\alpha*}})^2 \\
\sigma_{\delta,\text{total}}^2
&= \sigma_{\delta}^2 +
(\Delta{t}\, \sigma_{\mu_{\delta}})^2
\end{aligned}\right.,
\label{eq:exp:error}
\end{equation}
where $\sigma_{\alpha}$, $\sigma_{\delta}$, $\sigma_{\mu_{\alpha*}}$, and $\sigma_{\mu_{\delta}}$ are the uncertainties of the right ascension, declination, proper motion along right ascension, and proper motion along declination, respectively. $\Delta{T}$ is the difference between the catalog and the observation epochs. Since the epoch of the \gaia DR3 catalog is J2016.0, we set $\Delta{t} = \qty{13}{yr}$. The number of the reference sources was 10,888.

JASMINE aims at measuring the positions of the stars around the Galactic center with respect to the foreground reference stars. To imitate this circumstance, we added 1000 artificial sources with zero proper motions and zero parallaxes to the ground truth catalog. The artificial sources were randomly distributed within the area $(\ell, b) \in (\ang{-0.3}\pm\ang{0.3}, \ang{0.1}\pm\ang{0.3})$. The \texttt{source\_id}s of the artificial sources were manually assigned. The performance of the astrometric analysis was evaluated using the artificial sources.

\begin{table}
\centering
\caption{Columns of the Ground Truth Catalog / the Reference Catalog}
\label{tab:exp:catalog}
\begin{tabular}{>{\ttfamily}lcl}
\toprule
\multicolumn{1}{c}{Column name}
& \multicolumn{1}{c}{Symbol}
& \multicolumn{1}{c}{Comment} \\
\midrule
source\_id & $n$ & unique source ID \\
ra & $\alpha$ & right ascension \\
dec & $\delta$ & declination \\
ra\_error & $\sigma_{\alpha*}$ & uncertainty of right ascension, multiplied by $\cos\delta$ \\
dec\_error & $\sigma_\delta$ & uncertainty of declination \\
pmra & $\mu_{\alpha*}$ & proper motion in right ascension, multiplied by $\cos\delta$ \\
pmdec & $\mu_{\delta}$ & proper motion in declination \\
pmra\_error & $\sigma_{\mu_{\alpha*}}$ & uncertainty of proper motion in right acension \\
pmdec\_error & $\sigma_{\mu_{\delta}}$ & uncertainty of proper motion in declination \\
parallax & $\varpi$ & parallax \\
parallax\_error & $\sigma_\varpi$ & uncertainty of parallax \\
ref\_epoch & $t_0$ & reference epoch of the catalog \\
flag\_ref  & $f_\text{ref}$ & boolean flag for reference sources \\
flag\_test & $f_\text{test}$ & boolean flag for artificial sources \\
\bottomrule
\end{tabular}
\end{table}

\subsubsection{Reference catalog}
\label{sec:exp:survey:ref}

The astrometric analysis of JASMINE will rely on the reference information. We generated a reference catalog by resampling the ground truth catalog. How the astrometric parameters were sampled is summarized in \cref{tab:exp:resampling}. The uncertainties of the non-reference sources were replaced with the corresponding values. Here, we adopted large scatters for non-reference sources ($\qty{1}{arcsec}$ in positions) to pretend that we had almost zero astrometric information about non-reference sources.

\begin{table}[t]
\centering
\caption{Sampling Procedure}
\label{tab:exp:resampling}
\begin{tabular}{l@{\hspace{0.5cm}}c@{~}l@{\hspace{0.5cm}}c@{~}l}
\toprule
& \multicolumn{2}{c}{Reference Source}
& \multicolumn{2}{c}{Non-reference Source} \\
\midrule
Right Ascension
& $\hat{\alpha}$
& $\sim~ \mathcal{N}(\alpha, \sigma_{\alpha})$
& $\hat{\alpha}$
& $\sim~ \mathcal{N}(\alpha, \qty{1}{arcsec})$ \\
Declination
& $\hat{\delta}$
& $\sim~ \mathcal{N}(\delta, \sigma_{\delta})$
& $\hat{\delta}$
& $\sim~ \mathcal{N}(\delta, \qty{1}{arcsec})$ \\
Proper motion in RA
& $\hat{\mu}_{\alpha*}$
& $\sim~ \mathcal{N}(\mu_{\alpha*}, \sigma_{\mu_{\alpha*}})$
& $\hat{\mu}_{\alpha*}$
& $\sim~ \mathcal{N}(\mu_{\alpha*}, \qty{1}{\mas/yr})$ \\
Proper motion in Dec
& $\hat{\mu}_{\delta}$
& $\sim~ \mathcal{N}(\mu_{\delta}, \sigma_{\mu_\delta})$
& $\hat{\mu}_{\delta}$
& $\sim~ \mathcal{N}(\mu_{\delta}, \qty{1}{\mas/yr})$ \\
Parallax
& $\hat{\varpi}$
& $\sim~ \mathcal{N}(\varpi, \sigma_{\varpi})$
& $\hat{\varpi}$
& $\sim~ \mathcal{N}(\varpi, \qty{1}{\mas})$ \\
\bottomrule
\end{tabular}
\end{table}

\begin{figure}[t]
\centering
\includegraphics[width=0.95\linewidth]{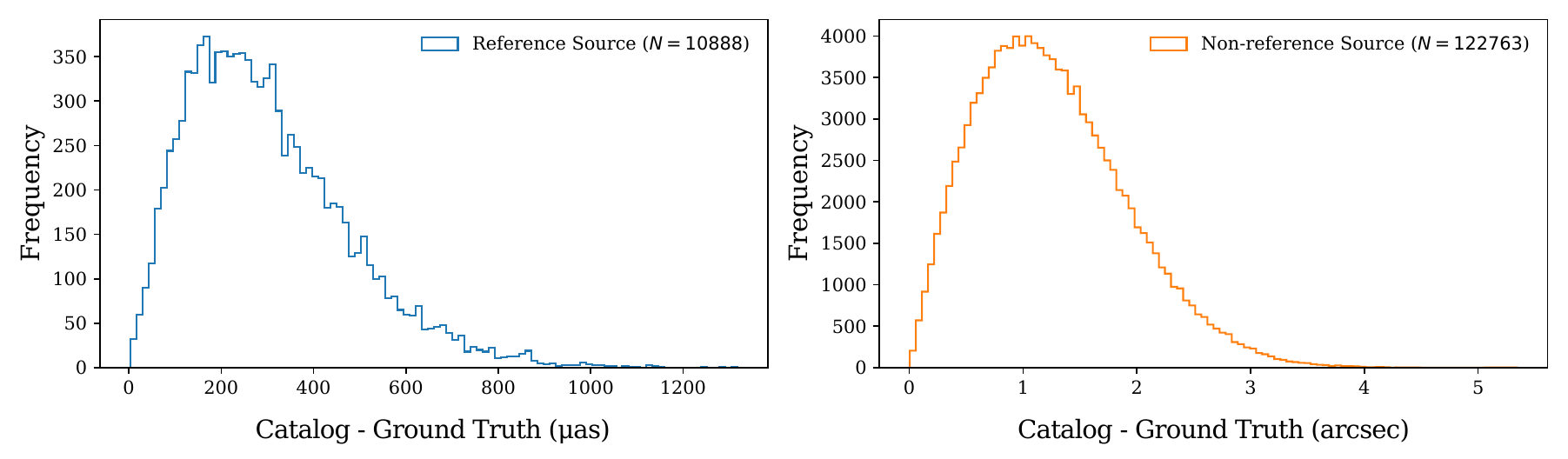}
\caption{Statistics of the on-sky separations between the positions calculated with the ground truth and reference catalogs at the observation epoch. The left and right panels illustrate the histograms for the reference and non-reference sources.}
\label{fig:exp:reference:hist}
\end{figure}

\cref{fig:exp:reference:hist} illustrates how the astrometric parameters were altered by resampling. The angular separations at the observation epoch (J2029.0) are evaluated. The left and right panels of \cref{fig:exp:reference:hist} show the histograms of the separations between the positions calculated at the observation epoch for the reference and non-reference sources, respectively. The separation angles for the reference sources range around \qtyrange{200}{500}{\uas}, chiefly due to the uncertainties of the proper motions. The separation angles for the non-reference sources are typically \qty{1.4}{arcsec}, corresponding to the positional uncertainties in resampling. This histogram suggests that the proper motions and parallaxes of the reference catalog are non-informative.

\subsubsection{Mock measurements}
\label{sec:exp:survey:mock}

Mock measurements were generated using the ground truth catalog and the observation schedule. For simplicity, we only included the aberration effect due to the Earth's orbital motion when calculating the stellar apparent positions. The positions on the detector were calculated using Eq.\,(\ref{eq:exp:model:nxny}). The mock measurements were sampled, assuming that the measurements followed the normal distributions.
\begin{equation}
n_{x,i,m} \sim \mathcal{N}(\hat{n}_{x,i,m}, \sigma^\text{obs}_{x,i,m}),
\quad
n_{y,i,m} \sim \mathcal{N}(\hat{n}_{y,i,m}, \sigma^\text{obs}_{y,i,m}).
\label{eq:exp:mock}
\end{equation}
Then, the likelihood part of \cref{eq:method:posterior:radec} is defined below.
\begin{equation}
\begin{aligned}
& \mathcal{P}(\mathcal{D}
~|~ \alpha_i^\text{src}, \delta_i^\text{src}, \ldots)
= \\
& \qquad
\prod_{i, m} ~
\mathcal{N}\left(n_{x,i,m} ~|~
\hat{n}_{x,i,m}(\alpha^\text{src}_{i}, \delta^\text{src}_{i},\ldots),
\sigma^\text{obs}_{x,i,m}\right)
\mathcal{N}\left(n_{y,i,m} ~|~
\hat{n}_{y,i,m}(\alpha^\text{src}_{i}, \delta^\text{src}_{i},\ldots),
\sigma^\text{obs}_{x,i,m}\right)
\end{aligned}
\label{eq:exp:likelihood}
\end{equation}
For the sake of simplicity, we assumed that the measurement uncertainties were common for all the sources. $\sigma^\text{obs}_{x, i}$ and $\sigma^\text{obs}_{y, i}$ were set to the values corresponding to \qty{4}{\mas} on the sky. The assumed measurement errors were about \qty{0.01}{pixel}, which were practically ultimate precision for wide-field camera systems.\cite{anderson_toward_2000} We adopted such high precision to assess possible systematic errors caused by the proposed method. A simulation study confirmed that the JASMINE optics possibly achieve \qty{4}{mas} precision in positional measurements, although the precision can be degraded due to the telescope jitter motion.\cite{kamizuka_jasmine_2024} The generated mock measurements are summarized in the ``Measurements'' section of \cref{tab:exp:parmaeters}.

\begin{table}[p]
\centering
\caption{Inputs, Model Parameters, and Priors}
\label{tab:exp:parmaeters}
\scalebox{0.9}{
\begin{tabular}{c>{\ttfamily}lcl}
\cmidrule[\heavyrulewidth]{2-4}
& \multicolumn{1}{c}{Name}
& \multicolumn{1}{c}{Symbol}
& \multicolumn{1}{c}{Explanation} \\
\cmidrule{2-4}
\multirow{9}{*}{\rotatebox{90}{\small Measurements}}
& orbit\_id  & $n$
& Orbit ID of the exposure ($n, \in [0, \ldots, 99]$) \\
& field\_id  & $m$
& Field ID of the exposure ($m, \in [0, 1, 2, 3]$) \\
& plate\_id  & $l$
& Exposure ID of the exposure ($l, \in [0, \ldots, 23]$) \\
& exptime & $t_{n,m,l}$
& timestamp of the exposure $(n,m,l)$ \\
& source\_id & $i$
& Source ID of the measurement \\
& nx & $n_{x,i,n,m,l}$
& $x$ pixel coordinate of the source $i$ in the exposure of $t_{n,m,l}$ \\
& sx & $\sigma_{x,i,n,m,l}$
& Uncertainties of $n_{x,i,n,m,l}$ \\
& ny & $n_{y,i,n,m,l}$
& $y$ pixel coordinate of the source $i$ at the exposure of $t_{n,m,l}$ \\
& sy & $\sigma_{y,i,n,m,l}$
& Uncertainties of $n_{y,i,n,m,l}$ \\
\cmidrule{2-4}
\multirow{16}{*}{\rotatebox{90}{\small Parameters}}
& ra & $\hat\alpha^\text{src}_{i}$
& right ascension of the source $i$ \\
& sig\_ra & $\sigma^\text{src}_{\alpha,i}$
& uncertainty of the right ascension of the source $i$ \\
& dec & $\hat\delta^\text{src}_{i}$
& declination of the source $i$ \\
& sig\_dec & $\sigma^\text{src}_{\delta,i}$
& uncertainty of the declination of the source $i$ \\
& ra\_tel & $\alpha^\text{tel}_{m}$
& right ascension of the telescope direction at the exposure $m$ \\
& dec\_tel & $\delta^\text{tel}_{m}$
& declination of the telescope direction at the exposure $m$ \\
& theta\_tel & $\theta^\text{tel}_{m}$
& position angle of the telescope pointing at the exposure $m$ \\
& foc\_tel & $F_{X}$
& focal plane scale at the exposure $m$ \\
& A\_kl & $A_{k,l}$
& focal plane distortion coefficients for the $X$ axis \\
& B\_kl & $B_{k,l}$
& focal plane distortion coefficients for the $Y$ axis \\
& det\_x & $X^\text{det}_n$
& focal plane $X$ coordinate of the detector $n$ \\
& det\_y & $Y^\text{det}_n$
& focal plane $Y$ coordinate of the detector $n$ \\
& det\_t & $\theta^\text{det}_n$
& focal plane rotation angle of the detector $n$ \\
& det\_sx & $\delta^\text{det}_{x,n}$
& focal plane $x$ pixel scale of the detector $n$ \\
& det\_sy & $\delta^\text{det}_{y,n}$
& focal plane $y$ pixel scale of the detector $n$ \\
\cmidrule{2-4}
\multirow{14}{*}{\rotatebox{90}{\small Priors}}
& ra
& $\mathcal{N}(\hat{\alpha}^\text{p}_{i}, \hat{\sigma}^\text{p}_{\alpha,i})$
& $\hat{\alpha}^\text{p}_{i}$ and $\hat{\sigma}^\text{p}_{\alpha,i}$
are obtained from the reference catalog \\
& dec
& $\mathcal{N}(\hat{\delta}^\text{p}_{i}, \hat{\sigma}^\text{p}_{\delta,i})$
& $\hat{\delta}^\text{p}_{i}$ and $\hat{\sigma}^\text{p}_{\delta,i}$ are
obtained from the reference catalog \\
& ra\_tel
& $\mathcal{N}(\hat{\alpha}^\text{tel}_{m}, \hat{\sigma}^\text{tel}_{\alpha,m})$
& $\hat{\alpha}^\text{tel}_{i,m}$ is obtained from the environment table;
$\hat{\sigma}^\text{tel}_{\alpha,m}$ is set to \ang{1} \\
& dec\_tel
& $\mathcal{N}(\hat{\delta}^\text{tel}_{m}, \hat{\sigma}^\text{tel}_{\delta,m})$
& $\hat{\delta}^\text{tel}_{i,m}$ is obtained from the environment table;
$\hat{\sigma}_{\delta,i,m}$ is set to \ang{1} \\
& theta\_tel
& \multicolumn{2}{l}{
Uniform distribution over $(\ang{-180}, \ang{180})$ }\\
& foc\_tel
& \multicolumn{2}{l}{
Gamma distribution with the mean of $\hat{F}_{m}$ and the variance of $100.0$
} \\
& A\_kl
& $\mathcal{N}(\hat{A}_{k,l}, \hat{\sigma}^A_{k,l})$
& $\hat{A}_{k,l} = 0$ and $\hat{\sigma}^A_{k,l} = 1.0$ \\
& B\_kl
& $\mathcal{N}(\hat{B}_{k,l}, \hat{\sigma}^B_{k,l})$;
& $\hat{B}_{k,l} = 0$ and $\hat{\sigma}^B_{k,l} = 1.0$ \\
& det\_x
& $\mathcal{N}(\hat{x}^\text{det}_n, \sigma^\text{det}_{x,n})$
& $\hat{X}^\text{det}_n$ is fixed to a rough estimate;
$\sigma^\text{det}_{x,n}$ is set to \qty{100.0}{\micro\meter} \\
& det\_y
& $\mathcal{N}(\hat{y}^\text{det}_n, \sigma^\text{det}_{y,n})$
& $\hat{Y}^\text{det}_n$ is fixed to a rough estimate;
$\sigma^\text{det}_{y,n}$ is set to \qty{100.0}{\micro\meter} \\
& det\_t
& $\mathcal{N}(\hat{\theta}^\text{det}_n, \sigma^\text{det}_{\theta,n})$
& $\hat{\theta}^\text{det}_n$ is fixed to a rough estimate;
$\sigma^\text{det}_{\theta,n}$ is set to \ang{0.1} \\
& det\_sx
& $\mathcal{N}(\hat{\delta}^\text{det}_{x,n}, \sigma^\text{det}_{\delta_x,n})$
& $\hat{\delta}^\text{det}_{x,n}$ is fixed to a rough estimate;
$\sigma^\text{det}_{\delta_x,n}$ is set to $0.01\hat{\delta}^\text{det}_{x,n}$ \\
& det\_sy
& $\mathcal{N}(\hat{\delta}^\text{det}_{y,n}, \sigma^\text{det}_{\delta_y,n})$
& $\hat{\delta}^\text{det}_{y,n}$ is fixed to a rough estimate;
$\sigma^\text{det}_{\delta_y,n}$ is set to $0.01\hat{\delta}^\text{det}_{y,n}$ \\
\cmidrule[\heavyrulewidth]{2-4}
\end{tabular}
}
\end{table}

\subsubsection{Parameter inference}
\label{sec:exp:survey:inference}

The procedure of the parameter inference is described in \cref{sec:method:optim}. The parameters of the observation model are listed in the ``Parameters'' section of \cref{tab:exp:parmaeters}. We assumed that the celestial coordinates followed the normal distributions, and that the probability density functions of the other parameters were given by delta functions for simplicity. The number of the \texttt{ra}, \texttt{sig\_ra}, \texttt{dec}, and \texttt{sig\_dec} parameters were as many as the number of unique sources, which was about 6,000--8,000 for a single orbit. Thus, the total number of the parameters ranged about 24,000--32,000.

The reference information was included as the priors. We calculated the estimated celestial positions and uncertainties at the observation using the reference catalog as follows.
\begin{align}
\hat{\alpha}^\text{p}_i
& = \alpha^\text{r}_{i,0}
+ \mu^\text{r}_{\alpha*,i}(t-t^\text{r}_0)
+ \varpi^\text{r}_{\alpha*,i}(t),
\label{eq:exp:survey:alpha_p} \\
\hat{\delta}^\text{p}_i
& = \delta^\text{r}_{i,0}
+ \mu^\text{r}_{\delta,i}(t-t^\text{r}_0)
+ \varpi^\text{r}_{\delta*,i}(t),
\label{eq:exp:survey:delta_p} \\
\bigl(\hat{\sigma}^\text{p}_{\alpha,i} \bigr)^2
& = \bigl(\hat{\sigma}^\text{r}_{\alpha,i} \bigr)^2
+ \bigl(\hat{\sigma}^\text{r}_{\mu_{\alpha*},i}(t-t^\text{r}_0)\bigr)^2
+ \bigl(\hat{\sigma}^\text{r}_{\varpi_\alpha,i} \bigr)^2,
\label{eq:exp:survey:sigma_alpha_p} \\
\bigl(\hat{\sigma}^\text{p}_{\delta,i} \bigr)^2
& = \bigl(\hat{\sigma}^\text{r}_{\delta,i} \bigr)^2
+ \bigl(\hat{\sigma}^\text{r}_{\mu_{\delta},i}(t-t^\text{r}_0)\bigr)^2
+ \bigl(\hat{\sigma}^\text{r}_{\varpi_\delta,i}\bigr)^2.
\label{eq:exp:survey:sigma_delta_p}
\end{align}
The terms with $r$ were derived from the reference catalogs. The $\varpi^\text{r}_{\alpha*,i}(t)$ and $\varpi^\text{r}_{\delta,i}$ were the parallaxes at the epoch $t$ along with the right ascension and declination, respectively. The priors of the celestial coordinates were provided by the normal distribution.
\begin{equation}
\mathcal{P}(
{\alpha}^\text{src}_{i}, {\delta}^\text{src}_{i},
\ldots
) = \prod_i
\mathcal{N}(
{\alpha}^\text{src}_{i} ~|~
\hat{\alpha}^\text{p}_i, \hat{\sigma}^\text{p}_{\alpha,i})
~
\mathcal{N}(
{\delta}^\text{src}_{i} ~|~
\hat{\delta}^\text{p}_i, \hat{\sigma}^\text{p}_{\delta,i}).
\label{eq:exp:priors}
\end{equation}
This formalism can handle the reference and non-reference stars similarly. Since the priors of the reference stars were tight, the celestial coordinates of the reference stars were strictly constrained. The uncertainties of the non-reference stars were large, and the priors were almost non-informative. The priors of the other parameters were manually assigned based on the assumption that rough estimates of the parameters were available.

The data analysis was conducted with the GPU cluster at the Center for Computational Astrophysics, National Astronomical Observatory of Japan. We used two computation nodes equipped with two AMD EPIC 7742 (64 cores) and eight NVIDIA A100 SXM (\qty{40}{GB}). The data obtained in a single orbit were calculated as a separate job. Each job was accelerated by a single GPU, and the jobs were processed in parallel by the GPU cluster. It took about 50--140 minutes to complete a single job.
\section{Results}
\label{sec:results}

\subsection{Plate Analysis}
\label{sec:result:pa}

\begin{figure}
\centering
\includegraphics[width=.8\linewidth]{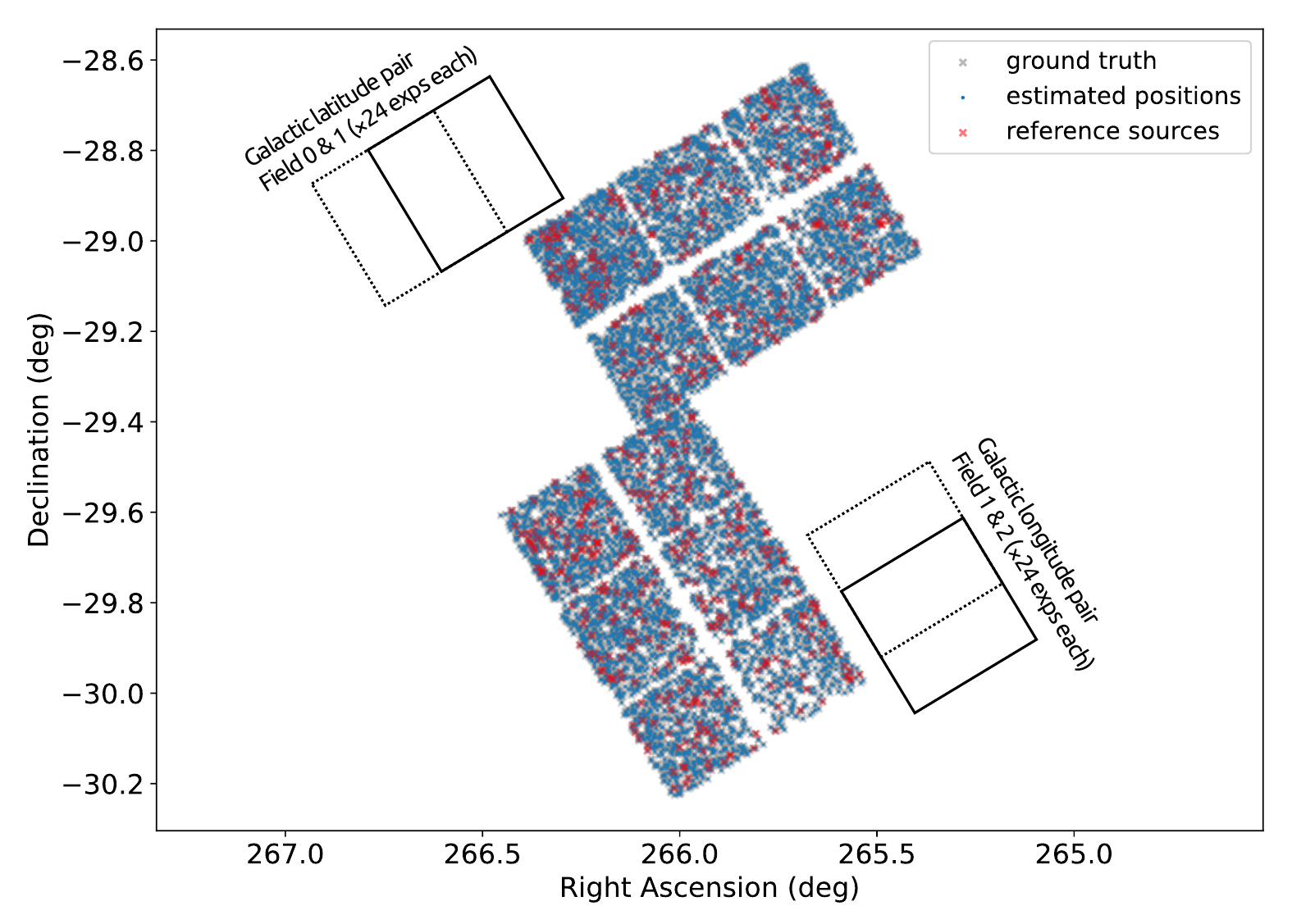}
\caption{Footprint of \textit{validation-20231120-0001}. The symbols show the distribution of the sources in the sky. The black rectangles indicate the alignment of the observation fields. In each field, 24 exposures are obtained.} The sources marked by the red crosses are the reference sources.
\label{fig:result:footprint}
\end{figure}

First, we present the result of \plateanalysis for each orbit using the data obtained in orbit 0001 processed on November 20, 2023 (\texttt{validation-20231120-0001}). \Cref{fig:result:footprint} illustrates the distribution of the sources in the dataset on the sky. The reference sources are marked in red.

The parameter convergence was checked based on the trace of the loss function and the normalized deviations of the reference sources. When the loss function did not change for more than $10^4$ iterations, we concluded that the parameters had converged to a stable solution. The normalized deviations were defined as follows:
\begin{equation}
\sigma^\text{ref}_{\alpha} = \left[
\frac{1}{N_\text{ref}}
\sum_{i \in \text{Ref}} \left(
\frac{\alpha^\text{src}_i - \hat{\alpha}^\text{p}_i}{\sigma^\text{p}_{\alpha, i}}
\right)^2
\right]^{\frac{1}{2}},
\quad
\sigma^\text{ref}_{\delta} = \left[
\frac{1}{N_\text{ref}}
\sum_{i \in \text{Ref}} \left(
\frac{\delta^\text{src}_i - \hat{\delta}^\text{p}_i}{\sigma^\text{p}_{\delta, i}}
\right)^2
\right]^{\frac{1}{2}},
\label{eq:result:deviation}
\end{equation}
where $N_\text{ref}$ is the total number of the reference sources. These values are expected to be unity when the parameters converge to the optimal point. We confirmed that the parameters were successfully optimized for all the orbits with respect to the criteria above.

\begin{figure}[p]
\centering
\includegraphics[width=0.8\linewidth]{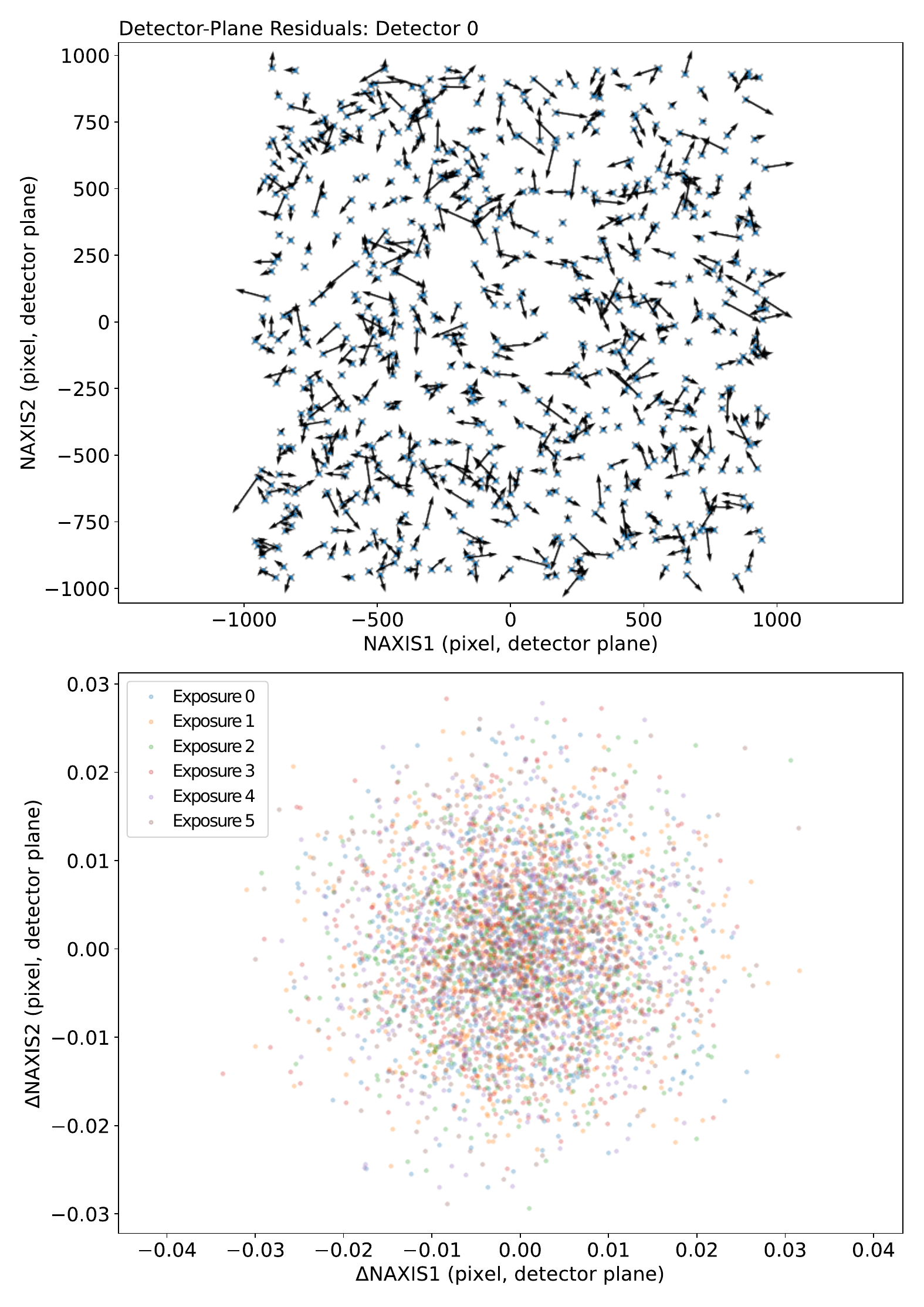}
\caption{Residuals between the measurements and the model predictions on the detector. The top panel shows a residual vector map on the detector \#0 for the exposure \#0. The bottom panel shows the residuals on the detector for the exposures \#0--\#5.}
\label{fig:result:residual:detector}
\end{figure}

The residuals on the detector, $\hat{n}_{x,i} - n_{x,i}$, are illustrated in \cref{fig:result:residual:detector}. The top panel shows the residuals of the first exposure in the first field as a vector map. The lengths of the vectors are enlarged for visibility. The directions of the vectors were randomly distributed, suggesting no global trend in the residuals. The bottom panel illustrates the scatter plot of the residuals for the first five exposures. The scatter of the residuals was about \qty{0.01}{pixel}, corresponding to about \qty{4}{\mas} on the sky, consistent with the measurement errors.

\begin{figure}[p]
\centering
\includegraphics[width=0.8\linewidth]{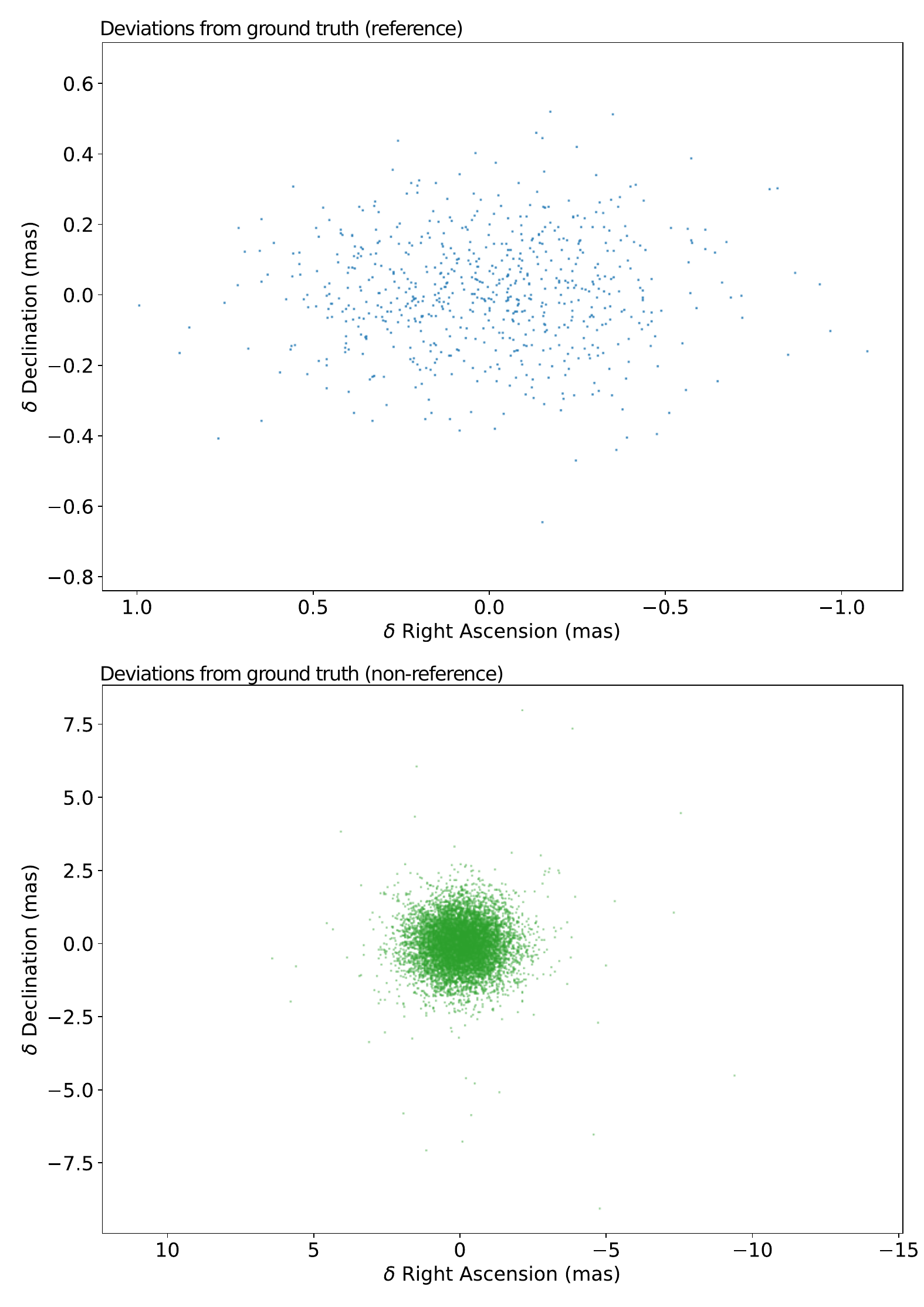}
\caption{Scatter plots of the residuals. The top and bottom panels show the residuals for the reference and non-reference stars, respectively.}
\label{fig:result:residual:onsky}
\end{figure}

\cref{fig:result:residual:onsky} illustrates the deviations of the estimated celestial coordinates from the ground truth. The top and bottom panels show the deviations for the reference and non-reference sources, respectively. The distribution for the reference sources was slightly elongated toward the right ascension. The size and shape of the distributions were consistent with the uncertainties of the priors. On the other hand, the deviations of the non-reference sources were circularly distributed with the size of about \qty{1}{\mas}, consistent with the measurement errors.

\begin{figure}[t]
\centering
\includegraphics[width=.9\linewidth]{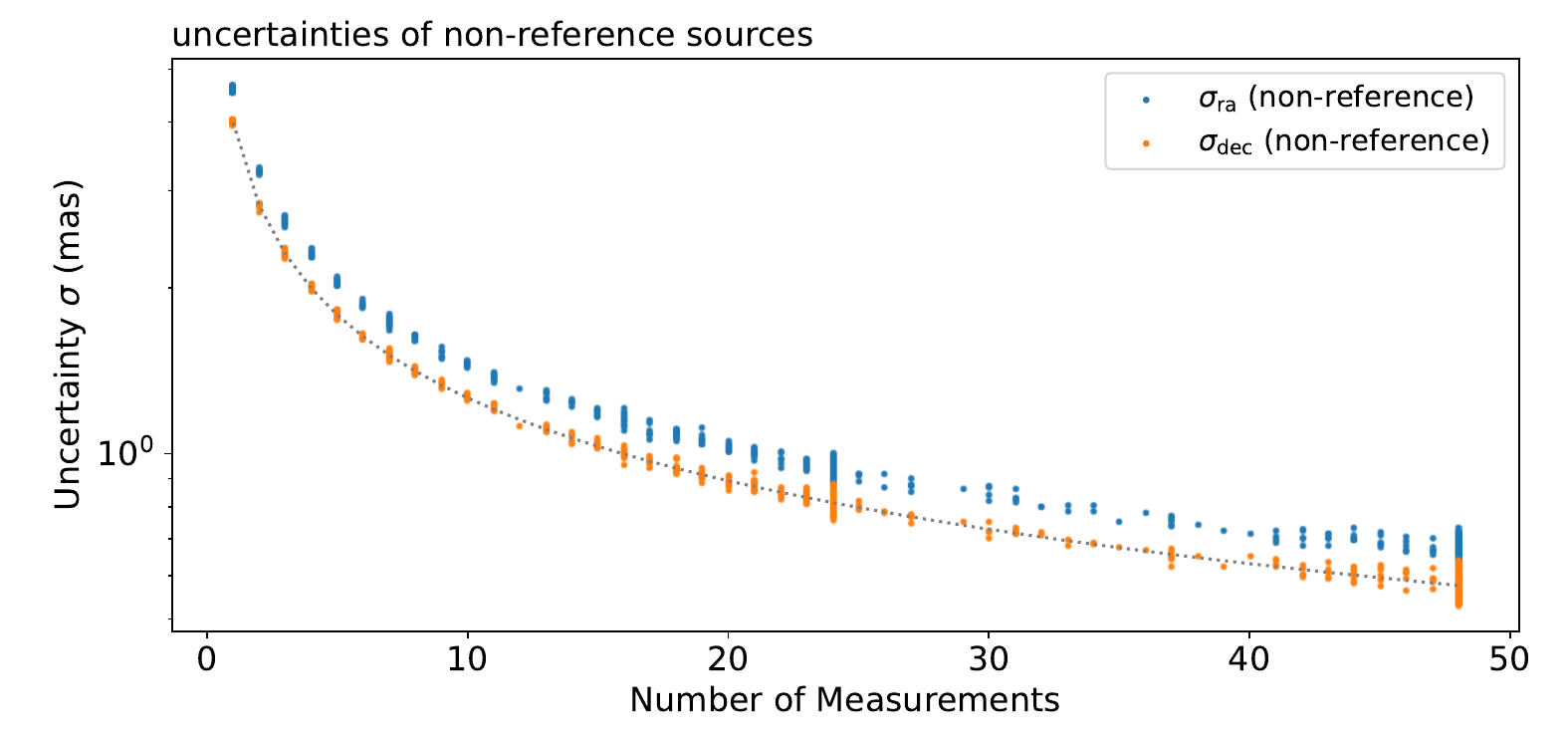}
\caption{Estimated uncertainties ($\sigma^\text{src}_{\alpha,i}, \sigma^\text{src}_{\delta,i}$) against the number of measurements. The gray dotted line indicates a $\sqrt{N}$-law assuming $\sigma_\text{exp} = \qty{4}{\mas}$ for reference.}
\label{fig:result:uncertainty}
\end{figure}

The number of measurements differed with sources since the telescope pointing jittered. The uncertainties of the estimated celestial coordinates were different accordingly. \Cref{fig:result:uncertainty} shows the estimated uncertainties against the number of measurements. The uncertainties followed a sqrt-$N$ low, $\sigma = \qty{4}{\mas} / \sqrt{N}$, indicated by the gray dotted line.

The following is a summary of the present results. The observation model converged to a stable solution using SVI with ADAM. No peculiar pattern was observed in the residuals on the detector plane, and the measurement errors well explained the magnitude of the residuals. The deviations from the ground truth were evaluated on the sky. The reference sources were consistent with the uncertainties of the priors, while the non-reference sources were consistent with the measurement errors. From these circumstances, we concluded that the parameters converged to an appropriate solution via \plateanalysis.

\subsection{Time-series astrometric analysis}
\label{sec:result:ts}

\begin{figure}[t]
\centering
\includegraphics[width=.9\linewidth]{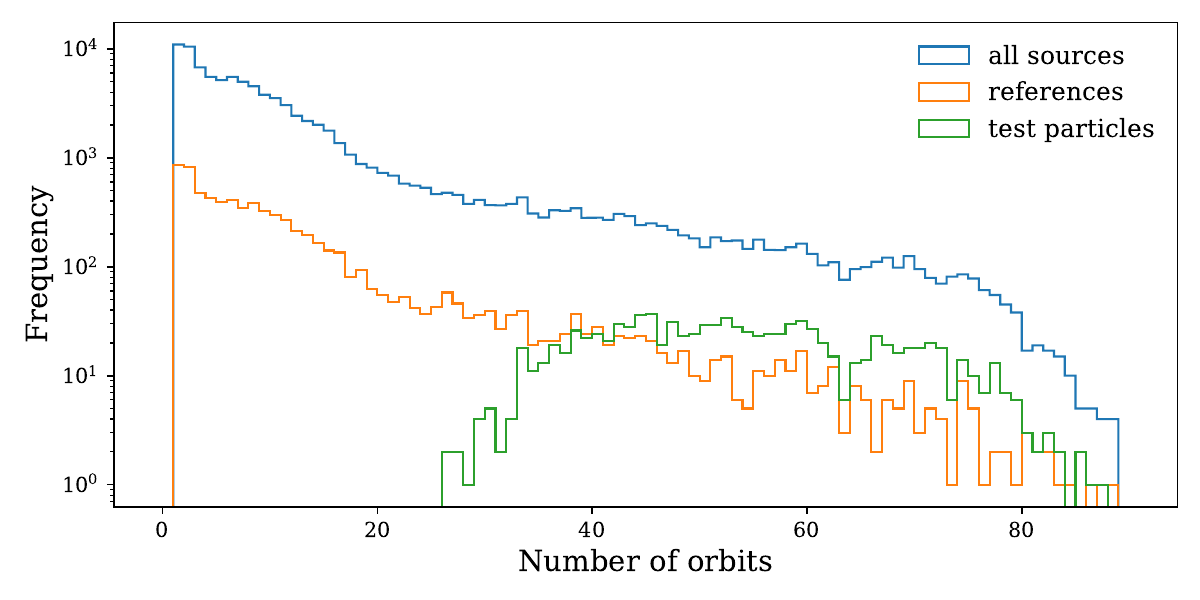}
\caption{Distributions of the number of orbits for all the sources (blue), the reference sources (orange), and the test particles (green). The sources around the target region are observed in more than 40 orbits.}
\label{fig:result:ts:orbit}
\end{figure}

We investigate the time series of the stellar celestial coordinates obtained by \plateanalysis. \Cref{fig:result:ts:orbit} summarizes the number of orbits where a source was observed in the JASMINE mini-mock survey. Sources around the target coordinate $(\ell, b) = (\ang{-0.3}, \ang{0.1})$ were typically observed for $>40$ orbits. The distributions of the reference and non-reference stars were similar, suggesting that two classes of sources were observed similarly. All the artificial sources were observed for more than 30 orbits since they were placed near the target coordinates.

In the following sections, we pick up the sources observed for more than 80 orbits and look into their motions in the sky with respect to the ground truth.

\subsubsection{Reference sources}
\label{sec:result:ts:ref}

\begin{figure}[tp]
\centering
\includegraphics[width=0.95\linewidth]{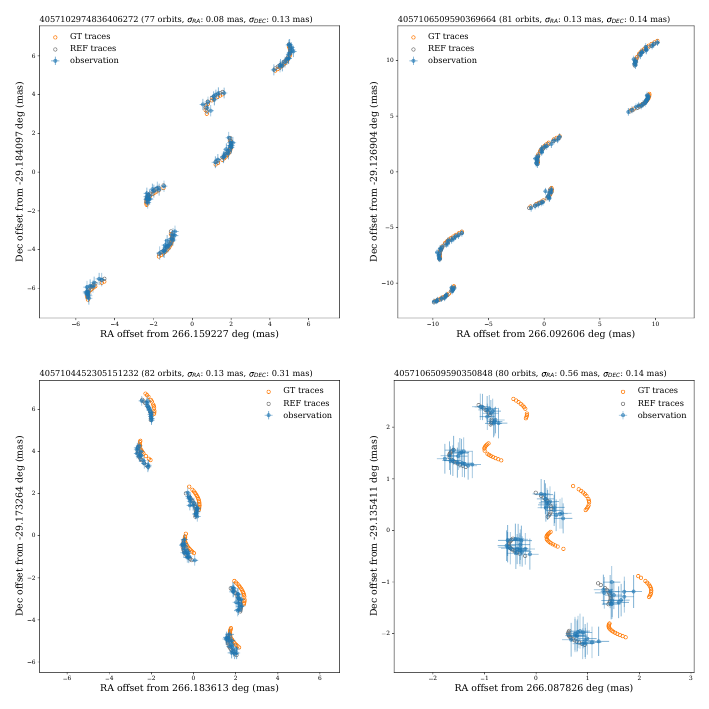}
\caption{On-sky motions of reference stars. The traces of the estimated coordinates are shown by the blue symbols with the error bars ($1\sigma$). The gray symbols show the apparent positions (proper motion + parallax) expected from the reference catalog, while the orange symbols illustrate the ground truth. The source ID and the number of orbits are annotated on the top with the root-mean-square separations from the ground truth. The large global offsets seen in the bottom panels are attributable to large prior errors.}
\label{fig:result:ts:ref}
\end{figure}

\Cref{fig:result:ts:ref} shows the on-sky motions of four representative reference stars. The coordinates estimated in the \textit{Plate Analysis} are shown by the blue symbols with errors. The orange symbols indicate the ground truth (apparent positions) of the sources. The gray symbols illustrate the coordinates expected from the reference catalog.

The panels of \cref{fig:result:ts:ref} suggest that the estimated coordinates well trace the ground truth. The root-mean-square separations from the ground truth were about \qtyrange{100}{500}{\uas}, while the standard deviations of the separations were about \qty{30}{\uas}. The priors tightly constrained the coordinates of the reference sources. Thus, large offsets were observed when the priors had significant errors (see the bottom panels of \cref{fig:result:ts:ref}). Their coordinates well reproduced the prior motions. The deviations between the priors and the ground truth are chiefly attributable to the propagation of the proper motion errors. The short-time motions (proper motions and parallaxes) are well reproduced, although the bottom panels show the global offsets.

\subsubsection{Non-reference sources}
\label{sec:result:ts:nonref}

\begin{figure}[tp]
\centering
\includegraphics[width=0.95\linewidth]{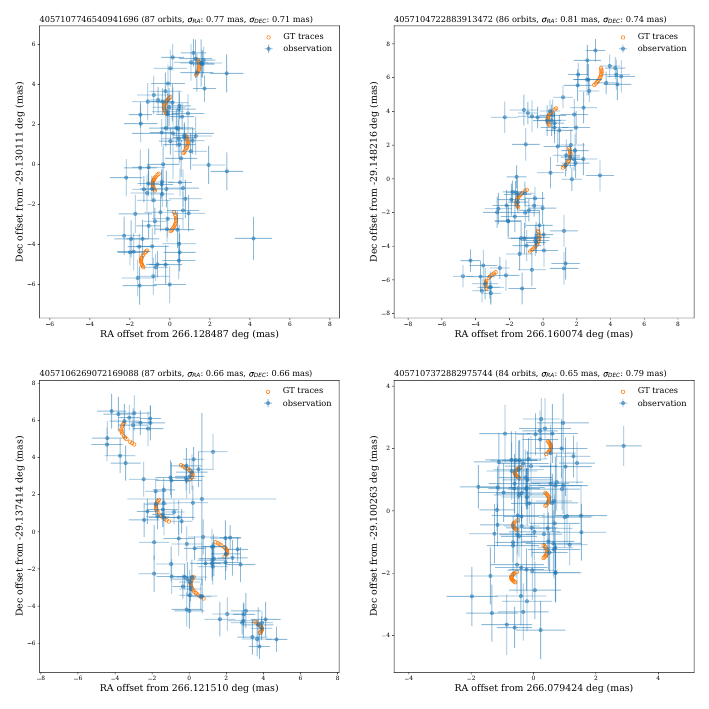}
\caption{On-sky motions of non-reference stars. The symbols are the same as in \cref{fig:result:ts:ref}, but the positions from the reference catalog are omitted.}
\label{fig:result:ts:nonref}
\end{figure}

The on-sky motions of four representative non-reference sources are presented in \cref{fig:result:ts:nonref}. The definitions of the symbols are the same as in \cref{fig:result:ts:ref}, but the coordinates expected from the reference catalog are removed since they appear far beyond the display range. The priors barely constrained these sources, but the estimated coordinates well traced the ground truth with some outliers. The root-mean-square separations ranged typically \qtyrange{600}{800}{\uas}.

Large systematic offsets like in the bottom right panel of \cref{fig:result:ts:ref} are not observed, possibly because the priors barely constrain the non-reference sources. However, the absolute coordinates of the non-reference sources indirectly depended on the priors of nearby reference sources. Systematic offsets in the celestial coordinates could be observed for a non-reference source close to a reference source with significant prior errors.

\subsubsection{Artificial sources}
\label{sec:result:ts:test}

\begin{figure}[tp]
\centering
\includegraphics[width=0.95\linewidth]{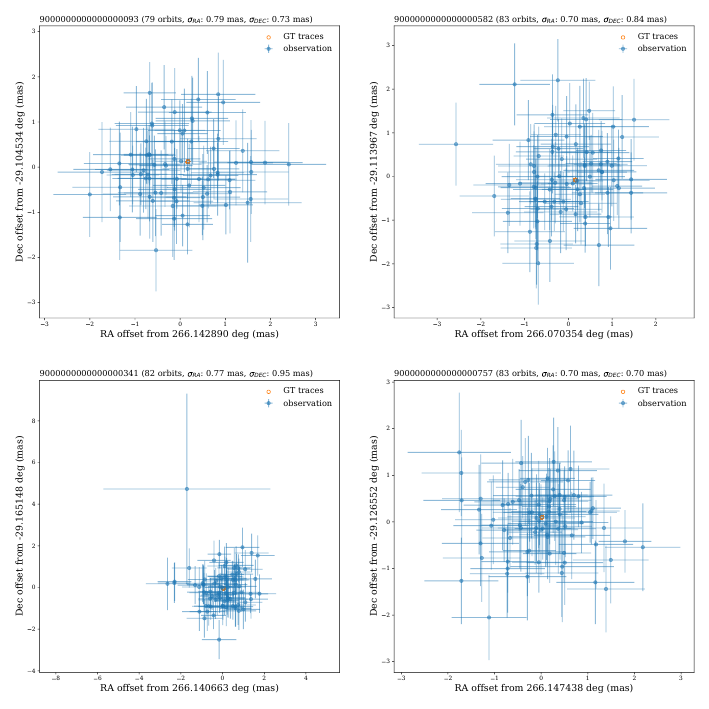}
\caption{On-sky motions of artificial sources. The symbols are the same as in \cref{fig:result:ts:ref}, but the positions from the reference catalog are omitted.}
\label{fig:result:ts:test}
\end{figure}

\cref{fig:result:ts:test} shows the distribution of the estimated coordinates for four representative artificial sources. The definitions of the symbols are the same as in \cref{fig:result:ts:nonref}. The ground truth symbols are located at the origins since the artificial sources have zero proper motions and zero parallaxes. The estimated coordinates are scattered around the origins within about $\qty{750}{\uas}$ in the root-mean-square measure, suggesting that the estimated coordinates were not significantly affected by surrounding sources.

\section{Discussion}
\label{sec:discussion}

We successfully obtain an astrometric solution of the observation model with about 30,000 parameters. The celestial coordinates and their uncertainties are simultaneously derived using the stochastic variational inference. The stochastic variational inference method with differentiable programming significantly accelerated the analysis. We found that the elapsed time depended linearly on the total number of measurements in the dataset, suggesting that the memory I/O limited the performance. In the actual mission, the number of parameters can be increased by a factor of three, and the number of exposures per field can be doubled. Although the computation cost may increase about six times from the JASMINE mini-mock survey, we presume that the \plateanalysis employed in the manuscript is still applicable to the actual mission.

The root-mean-square separations for the reference sources are about \qtyrange{100}{500}{\uas}, consistent with the positional uncertainties of the priors at the observation epoch shown in \cref{fig:exp:reference:hist}. However, the standard deviations of the separations are about \qty{30}{\uas}. Thus, the short-time motions (proper motions and parallaxes) were reproduced more precisely than the positional uncertainties of the priors. We presume that the errors were statistically mitigated since each exposure contained about 100 reference sources. The estimated positional uncertainties are about \qty{100}{\uas}. Such large uncertainties may reflect the inconsistency between the data and the priors. The results of the reference sources remain to be quantitatively investigated.

The root-mean-square separations for the non-reference sources are about \qty{0.7}{\mas}. Since each source is measured 36 times on average per orbit, a typical positional error for each orbit is estimated by $\sigma^\text{obs}/6 \simeq \qty{0.66}{\mas}$, consistent with the present results.

\begin{figure}[t]
\centering
\includegraphics[width=.9\linewidth]{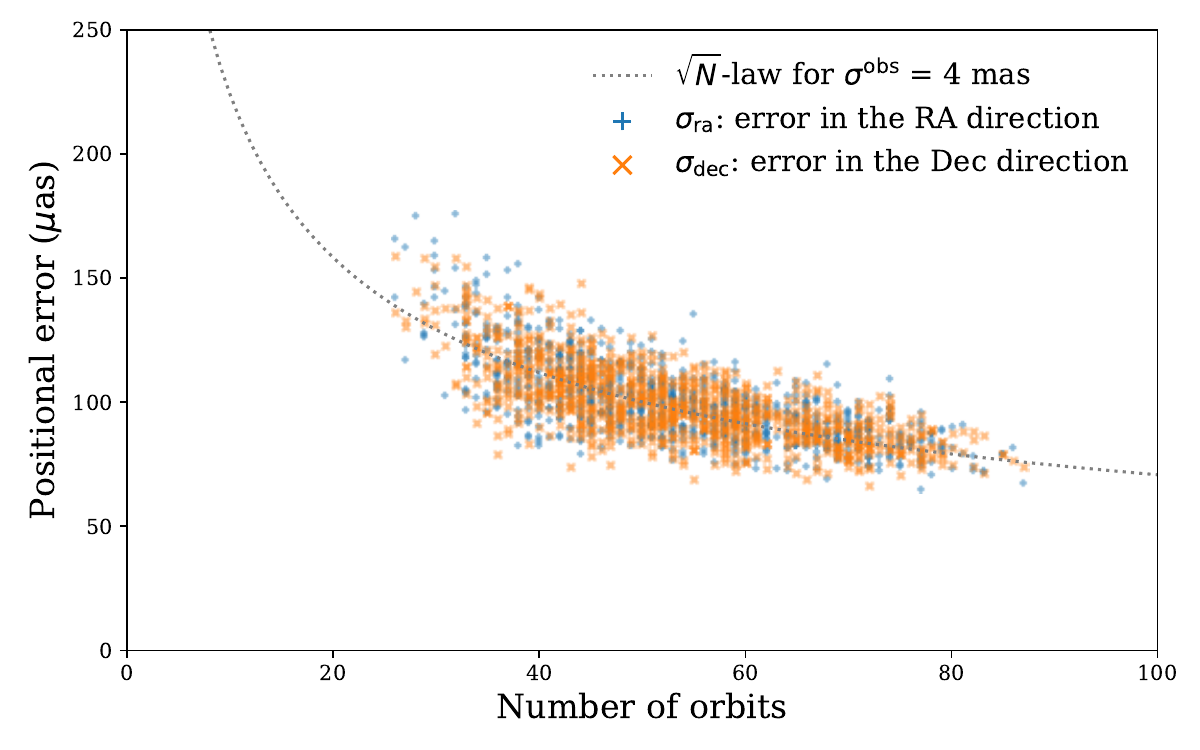}
\caption{Positional errors of the artificial sources estimated over the entire simulation. A $\sqrt{N}$-law for $\sigma^\text{obs} = \qty{4}{\mas}$ is illustrated by the gray dotted line for reference.}
\label{fig:discussion:test:rms}
\end{figure}

Similar statistics are evaluated for the artificial sources. We evaluate the positional errors based on the entire simulation and plot them against the number of orbits in \cref{fig:discussion:test:rms}. The gray dotted line shows the $\sqrt{N}$-law defined by $\varepsilon = \sigma^\text{obs}/\sqrt{36 N_\text{orbit}}$ for $\sigma^\text{obs} = \qty{4}{\mas}$ as a guide. The standard errors distribute around the $\sqrt{N}$-law without any apparent deviations, suggesting that the parameters of the observation model converged to an appropriate solution and every disturbing noise was removed to a level of \qty{70}{\uas}.

We confirm that \plateanalysis successfully estimated stellar positions at observation epochs while the distortion patterns were estimated as well. The achieved accuracy may depend on the positional uncertainties of reference sources. The telescope pointing direction and the plate scale are constrained only by the reference sources. If the positional uncertainties of the reference sources are comparable to measurement uncertainties, the uncertainties of estimated positions will be enhanced. If the reference sources are sparsely distributed and the density of non-reference sources is small, an estimated distortion pattern will be significantly affected by wrong reference priors. In such cases, we are obliged to limit the degree of freedom in the observation model (e.g., the order of distortion $N_\mathcal{L}$).

The present results suggest that \plateanalysis is generally successful. However, the current simulation is highly simplified and ignores several noise sources. JASMINE will orbit in the LEO at an altitude of about \qty{600}{km}. The aberration due to the satellite orbital motion is about 1/4 of the aberration by the Earth's orbital motion and changes in a timescale of about \qty{50}{min}. We should include the aberration caused by the satellite motion. Again, the aberration change during the exposure can distort images at a \qty{300}{\uas} level. We should update the observation model to accommodate such noise sources. In this experiment, we assigned the same measurement errors to all the sources. Realistic measurement errors should depend on the brightness of the sources. The satellite attitude control error (jittering motion) may change the measurement errors exposure by exposure. Further, source densities around the Galactic center are quite high. Neighboring sources may introduce systematic errors to measured positions. We assumed that the distortion pattern does not change within an orbit. However, the actual distortion pattern can change to some extent. When the point-spread function varies within a field of view, additional distortion that can change exposure by exposure can be introduced.\cite{kamizuka_jasmine_2024} The applicability of \plateanalysis can be limited by the abovementioned effects. Those effects should be considered to conclude the feasibility of the mission. The \gaia DR3 catalog was used as the ground truth catalog. We plan to compile a mock infrared catalog with astrometric information\cite{ramos_building_2024} and use it as the ground truth catalog.
\section{Conclusion}
\label{sec:conclusion}

For wide-field, relative astrometry, we propose an algorithm to estimate the coordinates of sources at the observation epoch with geometric distortion corrected using only science frames. The proposed algorithm is called \plateanalysis. We assume that the stellar motions are negligible and that the distortion pattern is stable during the observation period. Each field is assigned to the reference frame using reference sources. The image distortion is estimated using field stars in a series of exposures.

With a Bayesian approach, the posterior probabilistic distribution function is given by the product of the likelihood and the priors. The reference information is naturally introduced as the priors. To facilitate the analysis, we utilize the stochastic variational inference, where the posterior is approximated by the multivariate normal distribution. The parameter inference is significantly accelerated thanks to \texttt{JAX} and \texttt{numpyro}.

The proposed algorithm was validated based on numerical experiments. We conducted a simplified and small-scale survey simulation (JASMINE mini-mock survey), whose observational strategy emulates the Galactic center astrometry survey of the JASMINE mission. The simulated survey consists of observations only for 100 satellite orbits in three years. We generated mock measurements and estimated the stellar coordinates at each orbit. Then, the time sequences of the celestial coordinates were obtained.

The estimated coordinates were compared with the ground truth. In general, the estimated coordinates well reproduced the proper motions and parallaxes. However, the coordinates of some reference sources showed large offsets from the ground truth since the reference sources are tightly constrained by the priors and susceptible to errors of the priors. For the other sources, the separations from the ground truth are consistent with the given measurement errors. The performance of the \plateanalysis was evaluated using artificial sources with zero proper motions and zero parallaxes. The positional errors estimated over the entire simulation were consistently explained by the measurement errors, suggesting that every noise was removed at a level of about \qty{70}{\uas}. The present results suggest that the motions of distant sources around the Galactic bulge can be measured using the motions of foreground stars as references.

We confirmed that the \plateanalysis successfully worked in a simplified case, indicating the broad availability of the proposed algorithm. However, the current simulation ignores several factors: the observer's velocity, realistic measurement errors, and stellar distribution in the infrared. We will address these factors in future work to achieve ultimate accuracy. The present results suggest the potential of \plateanalysis for precise astrometric applications.

\subsection*{Disclosures}
The authors declare that there are no financial interests, commercial affiliations, or other potential conflicts of interest that could have influenced the objectivity of this research or the writing of this paper.

\subsection*{Data, Materials, and Code Availability}
The codes used in the numerical experiments are publicly hosted in the GitHub \href{https://github.com/JASMINE-Mission/jasmine_warpfield}{repository}\cite{ohsawa_jasmine_warpfield_2023}, and the dataset used in the performance verification is publicly hosted in Zenodo (\href{https://zenodo.org/records/10403895}{Version 20240602})\cite{ohsawa_jasmine_2024}.

\subsection*{Acknowledgments}
The main results of this paper were based on the article published in the SPIE proceedings (Ohsawa et al., 2024)\cite{ohsawa_concept_2024}. Numerical computations were in part carried out on the GPU system at the Center for Computational Astrophysics, National Astronomical Observatory of Japan. This research is in part supported by the JASMINE mission team and the Institute of Space and Astronautical Science (ISAS), Japan Aerospace eXploration Agency (JAXA). This work has made use of data from the European Space Agency (ESA) mission {\it Gaia} (\url{https://www.cosmos.esa.int/gaia}), processed by the {\it Gaia} Data Processing and Analysis Consortium (DPAC, \url{https://www.cosmos.esa.int/web/gaia/dpac/consortium}). Funding for the DPAC has been provided by national institutions, in particular the institutions participating in the {\it Gaia} Multilateral Agreement. DK acknowledges MWGaiaDN, a Horizon Europe Marie Sk\l{}odowska-Curie Actions Doctoral Network funded under grant agreement no. 101072454 and also funded by UK Research and Innovation (EP/X031756/1). DK also acknowledges the UK's Science \& Technology Facilities Council (STFC grant ST/W001136/1). This research made use of Astropy, a community-developed core Python package for Astronomy (Astropy Collaboration, 2013, 2018)\cite{astropy_collaboration_astropy_2013,astropy_collaboration_astropy_2022}. Part of this research uses NumPy (Harris et al., 2020)\cite{harris_array_2020} and Scipy, an open-source software for mathematics, science, and engineering (Virtanen et al., 2020)\cite{virtanen_scipy_2020}. The parameter inference is based on the probabilistic programming framework Numpyro (Phan et al., 2019)\cite{phan_composable_2019} and the auto-differentiation framework JAX (Bradbury et al., 2018)\cite{bradbury_jax_2018}. Grammarly and ChatGPT are used for the language and grammar clean-up.

\vspace{2ex}\noindent\textbf{Ryou Ohsawa} is an assistant professor at the National Astronomical Observatory of Japan, where he contributes to the JASMINE project. He completed his Ph.D. in Science at the University of Tokyo in 2014, with prior degrees in astronomy from the same institution. His research primarily explores optical and infrared astronomy, including interstellar medium physics and planetary science. Dr. Ohsawa has contributed to understanding interstellar dust formation, mid-infrared observational techniques, and astrometry. He is a member of the Astronomical Society of Japan, the Japanese Society for Planetary Science, and the Japan Geoscience Union.

\end{document}